\documentclass[aps,pra,onecolumn,showpacs,superscriptaddress,floatfix]{revtex4}
\usepackage{graphicx,times,nicefrac,amsmath,amsfonts,amssymb,amsthm,epsf,bm,bbm,longtable,dcolumn}
\usepackage{revsymb,wasysym,mathrsfs,color}
\begin{document}

\title{Differential cross sections for K-shell ionization by electron or positron impact}

\author{A I Mikhailov}
\affiliation{Petersburg Nuclear Physics Institute, 188300 Gatchina,
St.~Petersburg, Russia}
\affiliation{ Max-Planck-Institut f\"ur
Physik komplexer Systeme, D-01187 Dresden, Germany}

\author{A V Nefiodov}
\affiliation{Petersburg Nuclear Physics Institute, 188300 Gatchina,
St.~Petersburg, Russia}
\affiliation{ Max-Planck-Institut f\"ur
Physik komplexer Systeme, D-01187 Dresden, Germany}

\author{G Plunien}
\affiliation{Institut f\"ur Theoretische Physik, Technische
Universit\"at Dresden, D-01062  Dresden, Germany}

\date{Received \today}
\widetext
\begin{abstract}
We have investigated the universal scaling behavior of differential
cross sections for the single K-shell ionization by electron or
positron impact. The study is performed within the framework of
non-relativistic perturbation theory, taking into account the
one-photon exchange diagrams. In the case of low-energy positron
scattering, the doubly differential cross section exhibits prominent
interference oscillations. The results obtained are valid for
arbitrary atomic targets with moderate values of nuclear charge
number $Z$. \\

\noindent (Some figures in this article are in colour only in the
electronic version)
\end{abstract}

\maketitle

\section{Introduction}

The single ionization of inner-shell electrons by lepton impact is
the fundamental atomic process of particular interest
\cite{1,2,3,4,5,6}. In the present paper, which is a further
extension of our previous works \cite{7,8}, we deduce the universal
scaling behavior of differential cross sections for the single
K-shell ionization of hydrogen-like multicharged ions by electron or
positron impact. Special emphasis is laid on the energy domain near
the ionization threshold, where accurate description of the
electron-electron and electron-nucleus interactions is extremely
significant. The study is performed to leading order of
non-relativistic perturbation theory with respect to the
electron-electron interaction. The nucleus of an ion is treated as a
source of the external field (Furry picture). Accordingly, the
Coulomb functions are employed as electron wave functions in a
zeroth approximation. Due to the universal scaling behavior of the
K-shell ionization cross sections, the results obtained allow for
generalization on the case of arbitrary non-relativistic atomic
targets. The parameter $\alpha Z$, where $\alpha$ is the
fine-structure constant, is supposed to be sufficiently small
($\alpha Z \ll 1$), while we assume that $Z \gg 1$. The relativistic
units are used throughout the paper ($\hbar=1$, $c=1$).

\section{Ionization of hydrogen-like ions}
\subsection{Electron impact}

Let us consider first the inelastic electron scattering on
hydrogen-like ion in the ground state, which results in ionization
of a K-shell bound electron. We shall derive formulas for the
differential cross sections of the process. An incident particle can
be characterized by the energy $E=\bm{p}^2/(2 m)$ and the momentum
$\bm{p}$ at asymptotically large distances from the nucleus. We
focus on the non-relativistic energies $E$ within the range $I
\lesssim E \ll m$, where $I=\eta^2/(2m)$ is the ionization
potential, $\eta =m \alpha Z$ is the average mo\-men\-tum of a K
shell electron, and $m$ is the electron mass.

\begin{figure}[b]
\centerline{\includegraphics[scale=0.6]{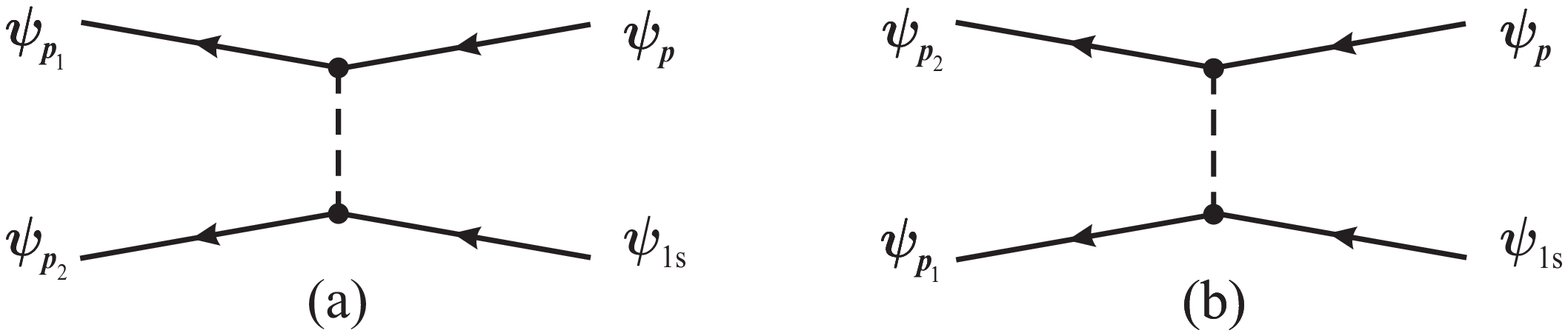}}
\caption{\label{fig1} Feynman diagrams for ionization of the K-shell
electron by an electron impact. Solid lines denote electrons in the
Coulomb field of the nucleus, while dashed line denotes the
electron-electron Coulomb interaction.}
\end{figure}

The process under consideration is described by the Feynman diagrams
depicted in Fig.~\ref{fig1}. In the final continuum state, the
electron wave functions are denoted as $\psi_{\bm{p}_1}$ and
$\psi_{\bm{p}_2}$. The energy conservation implies $E - I =E_1+E_2$,
where $E_1= \bm{p}_1^2/(2m)$ and $E_2 = \bm{p}_2^2/(2m)$ are the
energies of scattered and ejected electrons. In the single
ionization by low-energy particle impact, the emission of electrons
occurs with arbitrary energy sharing. In this case, both diagrams
depicted in Figs.~\ref{fig1}(a) and (b) give comparable
contributions to the ionization cross section and should be taken
into account. Accordingly, the total amplitude of the process reads
\begin{eqnarray}
{\mathcal A} &=& {\mathcal A}_a \delta_{\tau_1^\prime
\tau^{\phantom{\prime}}_1}\delta_{\tau_2^\prime
\tau^{\phantom{\prime}}_2}- {\mathcal A}_b \delta_{\tau_2^\prime
\tau^{\phantom{\prime}}_1}\delta_{\tau_1^\prime
\tau^{\phantom{\prime}}_2}\, , \label{eq1} \\
{\mathcal A}_a&=&\langle\psi_{\bm{p}_1}\psi_{\bm{p}_2} |
V_{\mathrm{C}}|\psi_{\bm{p}}\psi_{1s} \rangle\,  , \label{eq2} \\
{\mathcal A}_b&=&\langle\psi_{\bm{p}_2}\psi_{\bm{p}_1} |
V_{\mathrm{C}}|\psi_{\bm{p}}\psi_{1s} \rangle\,  . \label{eq3}
\end{eqnarray}
Here $\tau^{\phantom{\prime}}_{1,2}$ and $\tau_{1,2}^\prime$ denote
the spin projections of the Pauli spinors in the initial and final
states, respectively. The Coulomb interaction between two electrons
is described by the operator $V_{\mathrm{C}}(\bm{r}_1,\bm{r}_2) =
\alpha |\bm{r}_1- \bm{r}_2|^{-1}$, where $\bm{r}_1$ and $\bm{r}_2$
are the electron coordinates. The amplitude ${\mathcal A}_a$
corresponds to the direct diagram, while the amplitude ${\mathcal
A}_b$ is due to the exchange diagram.

As the wave functions of initial particles, we shall take the
corresponding solutions of the Schr\"{o}\-din\-ger equation for an
electron in the external field of the Coulomb source \cite{9}
\begin{eqnarray}
\psi_{1s}(\bm{r}) &=& N_{1s}e^{-\eta r}\, ,  \\
\psi_{\bm{p}}(\bm{r}) &=&\frac{4\pi}{2p}\sum_{l=0}^{\infty} i^l
e^{i\delta^{(-)}_{pl}} R^{(-)}_{pl}(r)\sum_{m=-l}^{l}
Y^{\phantom{*}}_{lm}\left(\hat{\bm{r}}\right)
Y^*_{lm}\left(\hat{\bm{p}}\right) \, . \label{eq5}
\end{eqnarray}
Here $N_{1s}^2 = \eta^3/\pi$, $\eta=m\alpha Z$ \cite{10},
$Y_{lm}(\hat{\bm{r}})$ are the spherical harmonics, which depend on
the variable $\hat{\bm{r}}=\bm{r}/r$, and $\delta^{(-)}_{pl}$ are
the phase shifts of the radial functions $R^{(-)}_{pl}$. The latter
are orthogonal and normalized according to
\begin{equation}
\int \limits_0^\infty dr r^2 R^{(-)}_{p'l}(r) R^{(-)}_{pl}(r) = 2\pi
\delta(p'-p) . \label{eq6}
\end{equation}
The asymptotical behavior of the wave function
$\psi_{\bm{p}}(\bm{r})$ is ``the sum of a plane wave and a
spherically outgoing wave". The functions \eqref{eq5} are normalized
by the condition
\begin{equation}
\int d\bm{r} \psi^*_{\bm{p}'}(\bm{r})
\psi^{\phantom{*}}_{\bm{p}}(\bm{r}) = (2\pi)^3\delta(\bm{p}'
-\bm{p}) \, . \label{eq7}
\end{equation}

For the Coulomb field of a point nucleus, one has \cite{9}
\begin{eqnarray}
R^{(\pm)}_{pl}(r) &=& \frac{C^{(\pm)}_{pl}}{(2l+1)!}(2 p r)^{l}e^{-i
p r } \Phi(l+1 \mp i\xi,2l+2,2ipr)\, ,  \label{eq8}\\
C^{(\pm)}_{pl}&=& 2p\, e^{\mp\pi\xi/2} |\Gamma(l+1 \pm i\xi)| \, , \label{eq9}\\
\delta^{(\pm)}_{pl}&=& \arg \Gamma(l+1 \pm i\xi) \,  , \label{eq10}
\end{eqnarray}
where $\xi=\eta/p$, $\Phi(x,y,z)$ is the confluent hypergeometric
function, and $\Gamma(z)$ is the Euler's gamma function. In
Eqs.~\eqref{eq8}--\eqref{eq10}, the lower (upper) sign corresponds
to the attraction (repulsion).

Let us assume that on experiment we are interested in the asymptotic
momentum of one outgoing electron only (for example, $\bm{p}_1$).
Then the wave function of this electron can be represented in terms
of the partial-wave decomposition
\begin{equation}
\psi_{\bm{p}^{\phantom{*}}_1}(\bm{r}) =\frac{4\pi}{2p_1}\sum_{l_1
,m_1} i^{l_1} e^{-i\delta^{(-)}_{p_1 l_1}} R^{(-)}_{p_1 l_1}(r)
Y_{l_1 m_1}\left(\hat{\bm{r}}\right)
Y^*_{l_1m_1}\left(\hat{\bm{p}}_1\right) \, , \label{eq11}
\end{equation}
which behaves asymptotically  as ``a plane wave plus a spherically
converging wave". The functions \eqref{eq11} are normalized by the
same condition \eqref{eq7}.

As a wave function of another outgoing electron, we shall take the
wave function of the stationary state characterized by the definite
values of the energy $E_2$, the angular momentum $l_2$, and its
projection $m_2$, namely,
\begin{equation}
\psi_{{p}^{\phantom{*}}_2}(\bm{r}) =R^{(-)}_{E_2l_2}(r)
Y_{l_2m_2}\left(\hat{\bm{r}}\right)\,  .
\end{equation}
The radial functions $R^{(\pm)}_{El}(r)$ are normalized to $\delta$
function in the energy
\begin{equation}
\int \limits_0^\infty dr r^2 R^{(\pm)}_{E'l}(r) R^{(\pm)}_{El}(r) =
\delta(E'-E) \, , \label{eq13}
\end{equation}
being related to $R^{(\pm)}_{pl}(r)$ as follows
\begin{equation}\label{eq14}
R^{(\pm)}_{El}(r)= \frac{\sqrt{m}}{\sqrt{2\pi p}}\,
R^{(\pm)}_{pl}(r)\, .
\end{equation}

The partial-wave decomposition of the Coulomb interaction
$V_{\mathrm{C}}$ is given by
\begin{equation}
V_{\mathrm{C}}(\bm{r}_1,\bm{r}_2)= \sum_{\lambda=0}^\infty
\frac{4\pi\alpha }{(2\lambda+1)}
\frac{r^\lambda_<}{r^{\lambda+1}_>}\sum_{\mu=-\lambda}^{\lambda}
Y^*_{\lambda \mu}\left(\hat{\bm{r}}_1\right)
Y^{\phantom{*}}_{\lambda\mu}\left(\hat{\bm{r}}_2\right) \,
,\label{eq15}
\end{equation}
where $r_< ={\rm min}\{r_1,r_2\}$ and $r_> ={\rm max}\{r_1,r_2\}$.

Choosing the $z$-axis along the momentum $\bm{p}$, we can perform
the integrations over the angular variables $\hat{\bm{r}}_1$ and
$\hat{\bm{r}}_2$ in the matrix elements \eqref{eq2} and \eqref{eq3}.
It yields
\begin{eqnarray}
{\mathcal A}_a&=&\frac{2\pi\alpha}{\eta^{2}}\frac{\sqrt{2\pi
m}}{\sqrt{p p_1}} \sum_{l,l_1} e^{i\Delta_{ll_1}}W^{l}_{l_1l_2}
C^{l0}_{l_1m_1 l_2m_2} Y^{\phantom{*}}_{l_1m_1}
\left(\hat{\bm{p}}_1\right)\,  ,  \label{eq16}\\
{\mathcal A}_b&=&\frac{2\pi\alpha}{\eta^{2}}\frac{\sqrt{2\pi
m}}{\sqrt{p p_1}} \sum_{l,l_1} e^{i\Delta_{ll_1}}V^{l}_{l_2l_1}
C^{l0}_{l_1m_1l_2m_2}Y^{\phantom{*}}_{l_1m_1}
\left(\hat{\bm{p}}_1\right)\,  , \label{eq17}\\
W^{l}_{l_1l_2}&=&\frac{1}{\sqrt{\pi k k_1 k_2}}
\frac{\Pi_{l_1}}{\Pi_{l_2}} C^{l0}_{l_1 0 l_2 0} I^l_{l_1l_2}\, ,
\label{eq18}\\
V^{l}_{l_2l_1}&=&\frac{1}{\sqrt{\pi k k_1 k_2}}
\frac{\Pi_{l_2}}{\Pi_{l_1}}C^{l0}_{l_1 0 l_2
0} J^l_{l_2l_1}\,  ,  \label{eq19}\\
I^l_{l_1l_2}&=& \int\limits_0^\infty dx^{\phantom{*}}_1 x_1^2
R^{(-)}_{k_1l_1}(x_1) R^{(-)}_{kl}(x_1) \int\limits_0^\infty
dx^{\phantom{*}}_2 x_2^2 R^{(-)}_{k_2l_2}(x_2)
\frac{x^{l_2}_<}{x^{l_2+1}_>} e^{-x_2}
\,  , \label{eq20}\\
J^l_{l_2l_1}&=& \int\limits_0^\infty dx^{\phantom{*}}_1
x_1^2R^{(-)}_{k_2l_2}(x_1)R^{(-)}_{kl}(x_1)\int\limits_0^\infty
dx^{\phantom{*}}_2 x_2^2 R^{(-)}_{k_1l_1}(x_2)
\frac{x^{l_1}_<}{x^{l_1+1}_>} e^{-x_2} \, , \label{eq21}
\end{eqnarray}
where $\Delta_{ll_1}= \delta^{(-)}_{pl}+\delta^{(-)}_{p_1l_1}+
\pi(l-l_1)/2$, $\Pi_{l}=\sqrt{2l+1}$, $x_< ={\rm min}\{x_1,x_2\}$,
$x_> ={\rm max}\{x_1,x_2\}$, and  $C^{lm}_{l_1 m_1 l_2 m_2}$ denotes
the Clebsch-Gordan coefficient. In Eqs.~\eqref{eq18}--\eqref{eq21},
we have introduced the dimensionless momenta $k=p/\eta$,
$k_i=p_i/\eta$ and the dimensionless coordinates $x_i=\eta r_i$,
$(i=1,2)$. Accordingly, the radial functions \eqref{eq8} satisfy to
the relation $R^{(-)}_{pl}(r) = \eta R^{(-)}_{kl}(x)$. Due to the
identity of electrons, the functions $V^{l}_{l_2l_1}$ can be
obtained from $W^{l}_{l_1l_2}$ by simultaneous substitutions $k_1
\rightleftharpoons k_2$ and $l_1 \rightleftharpoons l_2$.

The differential cross section for ionization of a K-shell electron
is related to the amplitude \eqref{eq1} as follows
\begin{equation}\label{eq22}
d\sigma^+_{\mathrm{K}} = \frac{2\pi}{v} \sum_{l_2,m_2} |{\cal A}|^2
\frac{d \bm{p}_1}{(2\pi)^3}\,dE_2  \delta(E_{1} + E_{2}+ I - E) \, ,
\end{equation}
where $v=p/m$ is the absolute magnitude of velocity of the incident
particle. The summations are performed over the angular momentum of
the second electron, because its state is not fixed. Equation
\eqref{eq22} defines distributions over the energy and scattering
angle. The element of the phase volume for electrons scattered into
the solid angle $d\Omega_1$ can be written as
\begin{equation}\label{eq23}
d\bm{p}_1 =  m p_1 \, dE_{1}\, d\Omega_1 \,  .
\end{equation}
In the case of unpolarized particles, Eq.~\eqref{eq22} should be
averaged over polarizations of the initial electrons and summed over
polarizations of the final electrons. This can be achieved by means
of the following substitution
\begin{equation}
|{\cal A}|^2 \to \overline{\vphantom{{\cal A}^2} | {\cal
A}|\,}\!{}^2 =\frac{1}{4}\sum_{\tau^{\vphantom{*}}_{1}, \tau_{1}'}
\sum_{\tau^{\vphantom{*}}_{2}, \tau_{2}'} |{\cal A}|^2 \, .
\label{eq24}
\end{equation}
Then the differential cross section takes the form
\begin{equation}
\frac{d\sigma^+_{\mathrm{K}}}{dE_1d\Omega_1} =
\frac{m^2}{(2\pi)^2}\frac{p_1}{p} \, \sum_{l_2,m_2}\Bigl\{
|{\mathcal A}_a|^2 + |{\mathcal A}_b|^2 - \frac{1}{2}\bigl(
{\mathcal A}^{\vphantom{*}}_a{\mathcal A}_b^* + {\mathcal
A}_a^*{\mathcal A}^{\vphantom{*}}_b \bigr) \Bigr\} \,   .
\label{eq25}
\end{equation}

Let us introduce the dimensionless energies $\varepsilon =E/I$ and
$\varepsilon_i=E_i/I$, $(i=1,2)$, where $I=\eta^2/(2m)$ is the
ionization potential for the K-shell electron. The
energy-conservation law now reads $\varepsilon -1 = \varepsilon_1
+\varepsilon_2$, where $\varepsilon=k^2$ and $\varepsilon_i =
k_i^2$, $(i=1,2)$. Inserting Eqs.~\eqref{eq16} and \eqref{eq17} into
Eq.~\eqref{eq25} yields
\begin{eqnarray}
\frac{d\sigma^+_{\mathrm{K}}}{d\varepsilon_1d \Omega_1} &=&
\frac{\sigma_0}{Z^4\varepsilon} \sum_{l',l_1'}
\sum_{l,l_1}\sum_{l_2,m_2}\cos
\bigl(\Delta^{l'l_1'}_{ll_1}\bigr)T^{l'l_1'}_{ll_1l_2}
C^{l0}_{l_1m_1 l_2m_2}C^{l'0}_{l_1'm_1 l_2m_2}
Y^{\phantom{*}}_{l_1m_1}\left(\hat{\bm{p}}_1\right)
Y^*_{l_1'm_1}\left(\hat{\bm{p}}_1\right) \,  ,  \label{eq26}\\
T^{l'l_1'}_{ll_1l_2}&=& W^{l}_{l^{\vphantom{'}}_1l^{\vphantom{'}}_2}
W^{l'}_{l_1'l^{\vphantom{'}}_2} +
V^{l}_{l^{\vphantom{'}}_2l^{\vphantom{'}}_1}
V^{l'}_{l^{\vphantom{'}}_2l_1'}- \frac{1}{2} \Bigl(
W^{l}_{l^{\vphantom{'}}_1l^{\vphantom{'}}_2}
V^{l'}_{l^{\vphantom{'}}_2l_1'} +
V^{l}_{l^{\vphantom{'}}_2l^{\vphantom{'}}_1}
W^{l'}_{l_1'l^{\vphantom{'}}_2} \Bigr)\,  . \label{eq27}
\end{eqnarray}
Here $\Delta^{l'l_1'}_{ll_1}= \Delta_{ll_1}- \Delta_{l'l_1'}$ and
$\sigma_0=\pi a_0^2=87.974$ Mb, where $a_0=1/(m\alpha)$ is the Bohr
radius.

In Eq.~\eqref{eq26}, the summation over the projection $m_2$ can be
easily performed using the following relation \cite{11}
\begin{eqnarray}
\sum_{m_2}C^{l0}_{l_1m_1 l_2m_2}C^{l'0}_{l_1'm_1 l_2m_2}
Y^{\phantom{*}}_{l_1m_1}\left(\hat{\bm{p}}_1\right)
Y^*_{l_1'm_1}\left(\hat{\bm{p}}_1\right) &=& \frac{1}{4\pi}
(-1)^{l_2}\Pi_{ll^{\phantom{*}}_1l'l_1'}\nonumber\\
&&\times \sum_{L\geqslant 0}P_L(\cos \theta_1) C^{L0}_{l 0
l'0}C^{L0}_{l^{\phantom{*}}_1 0 l_1'0} \Biggl\{
\begin{matrix} l_1 & l_2 & l \\ l' &L &l_1' \end{matrix}
\Biggr\}  \,  . \label{eq28}
\end{eqnarray}
Here $\Pi_{l l_1\ldots}= \sqrt{(2l+1)(2l_1+1)\cdots}$, $P_L(x)$
denotes the Legendre polynomial of order $L$, and the standard
notation for the $6j$-symbol is used. The scattering angle
$\theta_1$ is enclosed by the vectors $\bm{p}$ and $\bm{p}_1$. The
formula \eqref{eq28} possesses the axial symmetry with respect to
the direction $\bm{p}$ of incoming particles. Accordingly, the solid
angle $d\Omega_1$ is given by $d\Omega_1 =2 \pi \sin \theta_1
d\theta_1$.

Taking into account Eq.~\eqref{eq28}, the differential cross section
can be cast into the following form
\begin{eqnarray}
\frac{d\sigma^+_{\mathrm{K}}}{d\varepsilon_1 d\Omega_1} &=&
\frac{\sigma_0}{4\pi Z^4 }G(\varepsilon,\varepsilon_1,\theta_1) \, ,
\label{eq29}\\
G(\varepsilon,\varepsilon_1,\theta_1)&=&
F(\varepsilon,\varepsilon_1) + \sum_{L\geqslant
1}F_L(\varepsilon,\varepsilon_1) P_L(\cos \theta_1)
\,  , \label{eq30}\\
F(\varepsilon,\varepsilon_1) &=&\frac{1}{\varepsilon} \sum_{l, l_1,
l_2} T^{ll_1}_{ll_1 l_2}= \frac{1}{\varepsilon}\sum_{l, l_1,l_2}
\Bigl\{ \bigl(W^{l}_{l_1 l_2}\bigr)^2 + \bigl(V^{l}_{l_2
l_1}\bigr)^2 - W^{l}_{l_1 l_2} V^{l}_{l_2 l_1} \Bigr\} \,  , \label{eq31}\\
F_L(\varepsilon,\varepsilon_1) &=& \frac{1}{\varepsilon} \sum_{l,
l_1,l_2}\sum_{l',l_1'} (-1)^{l_2}\cos \bigl(\Delta^{l'l_1'}_{ll_1}
\bigr)T^{l'l_1'}_{ll_1l_2} \Pi_{ll^{\phantom{*}}_1l'l_1'} C^{L0}_{l
0 l'0}C^{L0}_{l^{\phantom{*}}_1 0 l_1'0} \Biggl\{
\begin{matrix} l_1 & l_2 & l \\ l' &L &l_1' \end{matrix}
\Biggr\} \,  . \label{eq32}
\end{eqnarray}
The angular dependence of the cross section \eqref{eq29} is governed
by the Legendre polynomials $P_L(\cos \theta_1)$. The functions
\eqref{eq30}--\eqref{eq32} are universal, being independent of the
nuclear charge $Z$. The energy $\varepsilon_1$ of outgoing electrons
lies within the range $0 \leqslant \varepsilon_1 \leqslant
\varepsilon -1$. The limiting values of $\varepsilon_1=0$ and
$\varepsilon_1= \varepsilon -1$ correspond to the situation, when
one of the electrons in the final state is infinitely slow. The
function \eqref{eq32} coincides with the function \eqref{eq31} in
the particular case, if $L=0$.

\begin{figure}[tbhp]
\begin{minipage}[t]{0.49\textwidth}
\centering\includegraphics[width=0.9\textwidth,angle=0,clip]{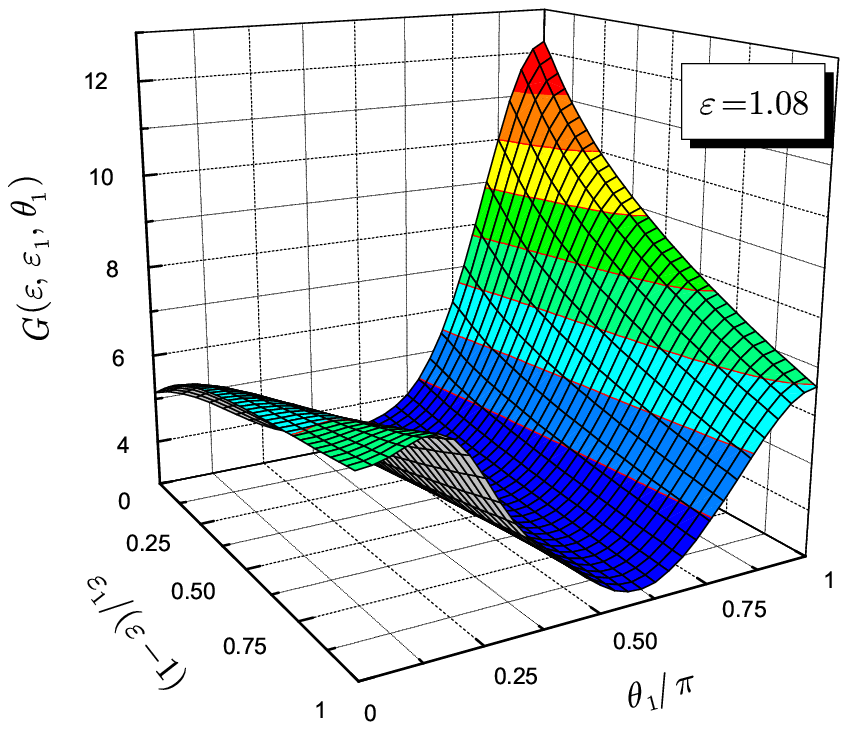}
\end{minipage}
\begin{minipage}[t]{0.49\textwidth}
\centering\includegraphics[width=0.9\textwidth,angle=0,clip]{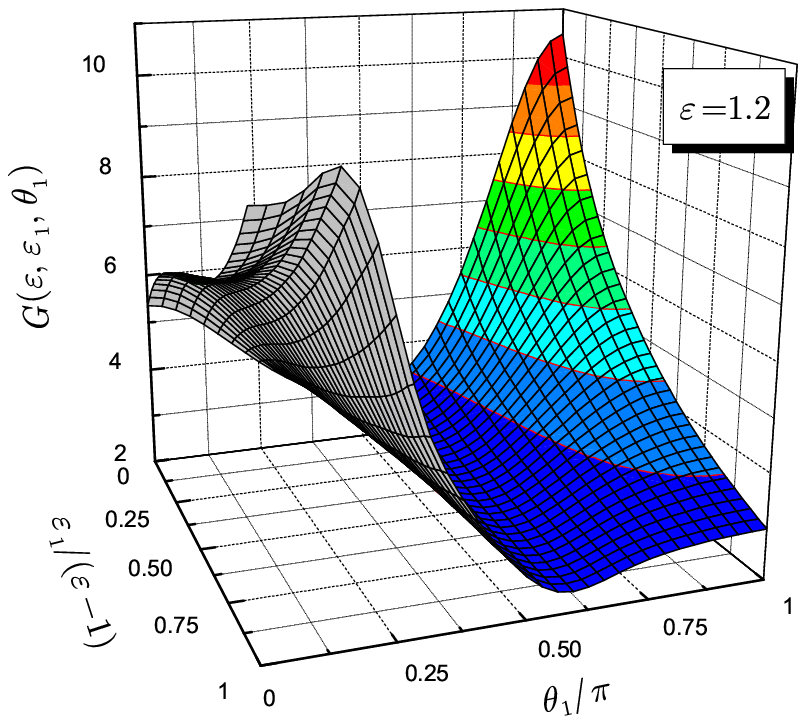}
\end{minipage}
\begin{minipage}[h]{0.49\textwidth}
\centering\includegraphics[width=0.9\textwidth,angle=0,clip]{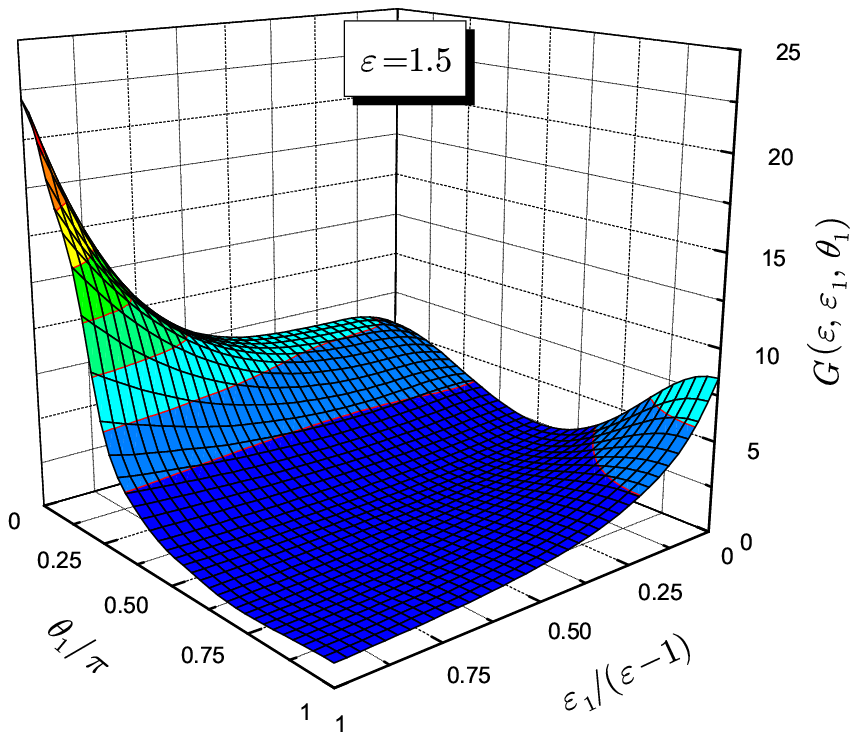}
\end{minipage}
\begin{minipage}[h]{0.49\textwidth}
\centering\includegraphics[width=0.9\textwidth,angle=0,clip]{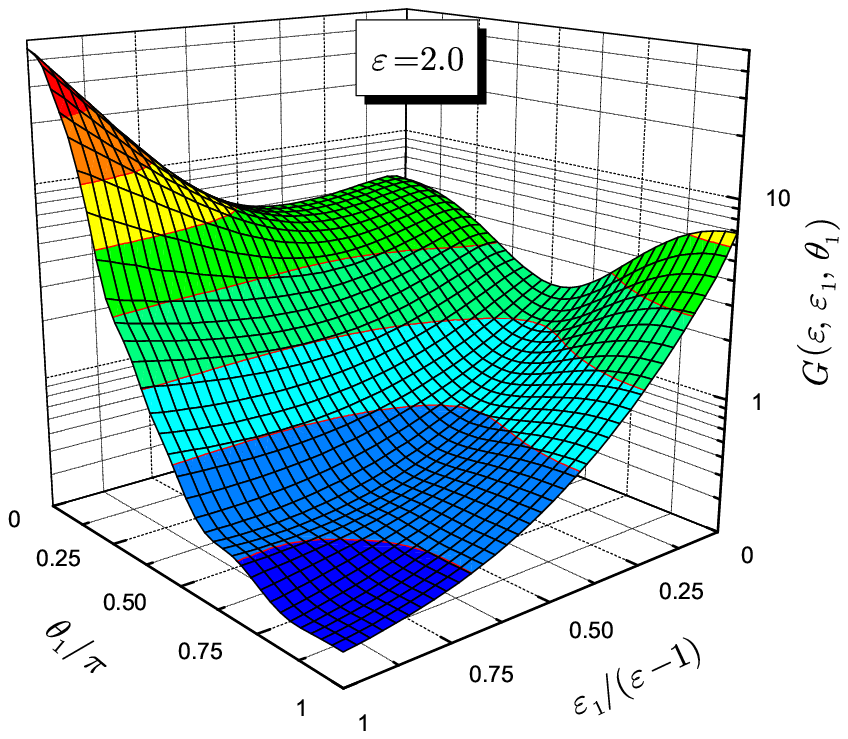}
\end{minipage}
\begin{minipage}[b]{0.49\textwidth}
\centering\includegraphics[width=0.9\textwidth,angle=0,clip]{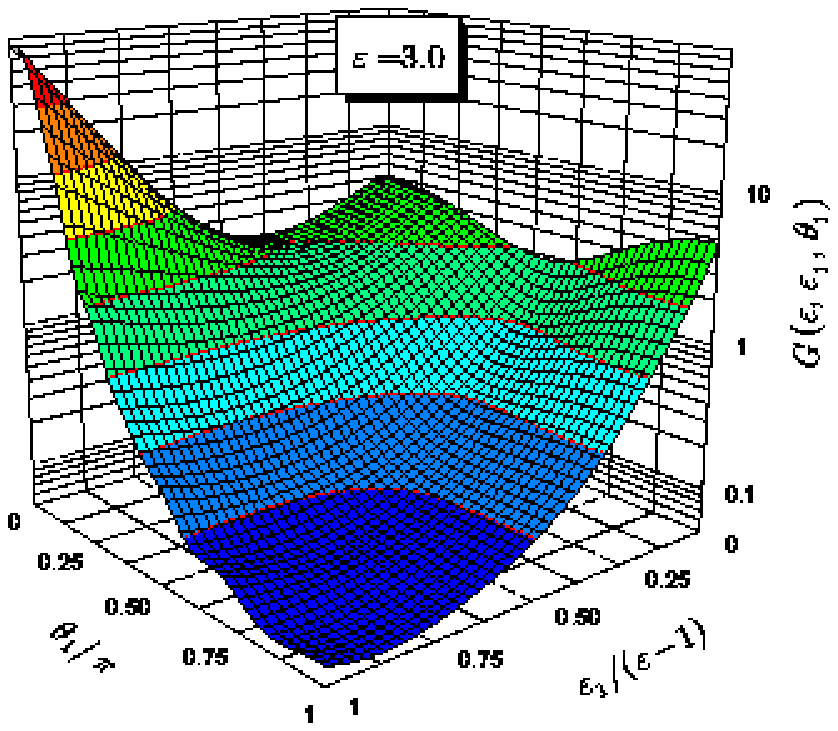}
\end{minipage}
\begin{minipage}[b]{0.49\textwidth}
\centering\includegraphics[width=0.9\textwidth,angle=0,clip]{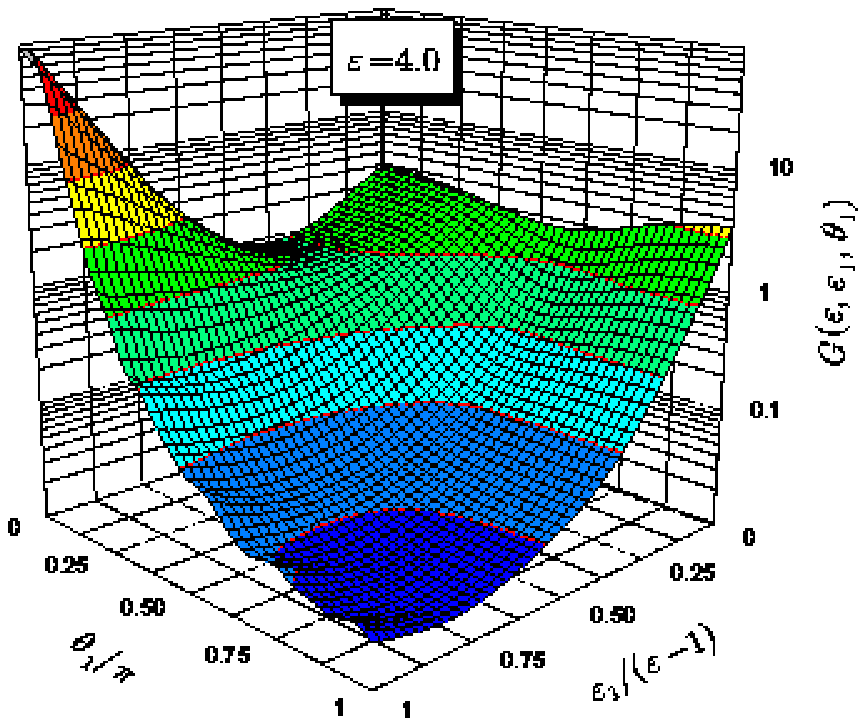}
\end{minipage}
\caption{\label{fig2} The universal function \eqref{eq30} is
calculated for different values of the dimensionless energy
$\varepsilon$ of the incident electron. The variable $\varepsilon_1$
is the energy of outgoing electron, which is detected at the angle
$\theta_1$. The center point $\varepsilon_1= (\varepsilon -1)/2$
corresponds to the equal-energy sharing $(\varepsilon_1
=\varepsilon_2)$.}
\end{figure}

\begin{figure}[tbhp]
\begin{minipage}[t]{0.49\textwidth}
\centering\includegraphics[width=0.9\textwidth,angle=0,clip]{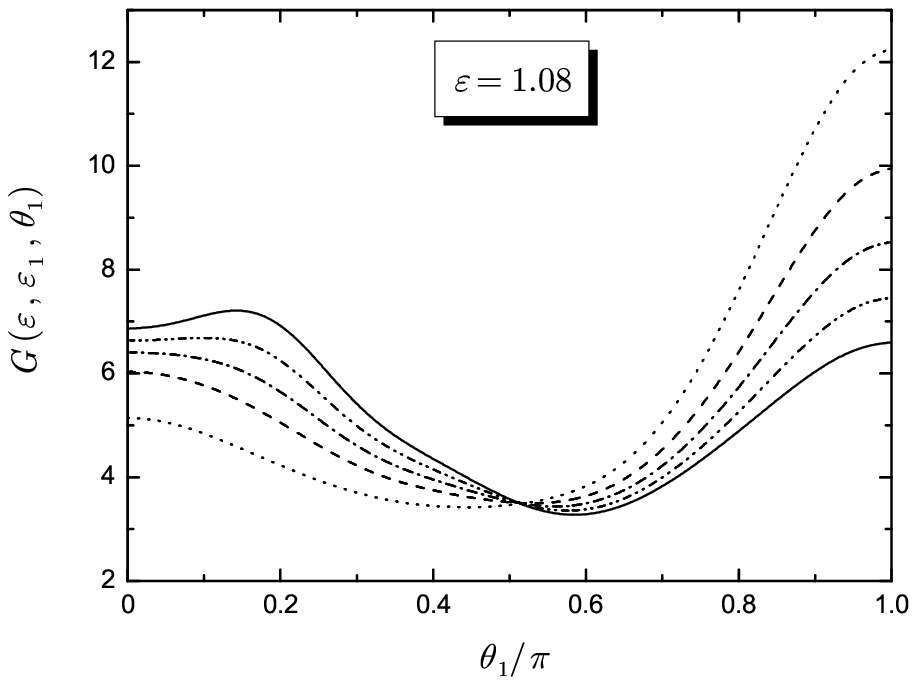}
\end{minipage}
\begin{minipage}[t]{0.49\textwidth}
\centering\includegraphics[width=0.9\textwidth,angle=0,clip]{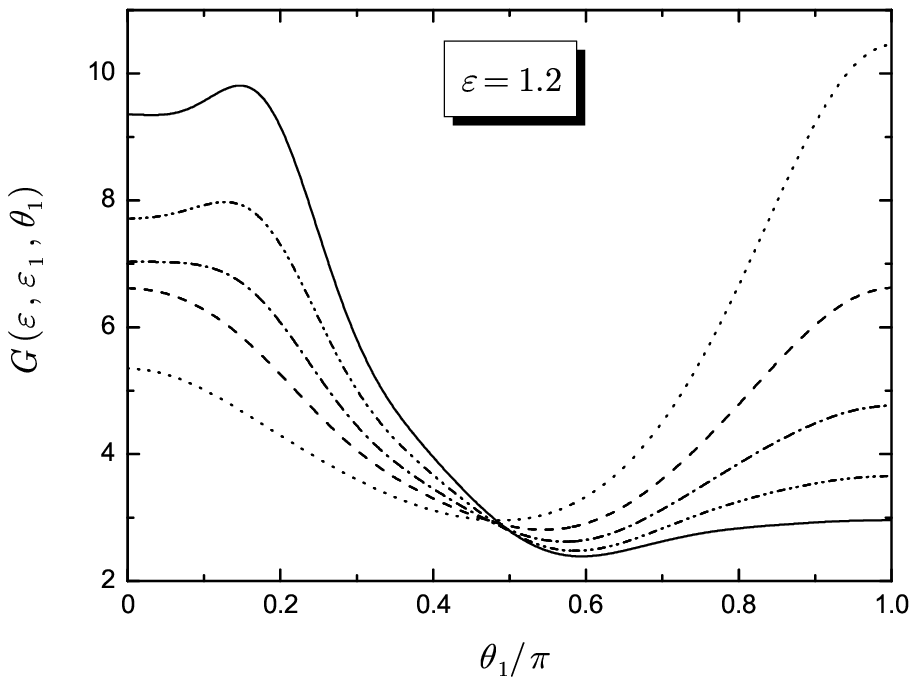}
\end{minipage}
\begin{minipage}[h]{0.49\textwidth}
\centering\includegraphics[width=0.9\textwidth,angle=0,clip]{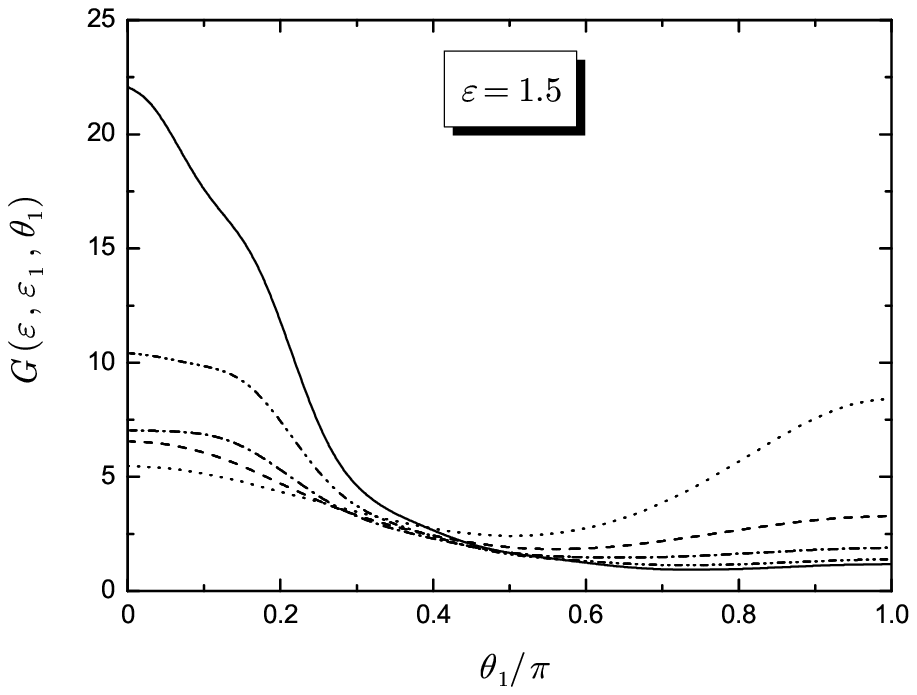}
\end{minipage}
\begin{minipage}[h]{0.49\textwidth}
\centering\includegraphics[width=0.9\textwidth,angle=0,clip]{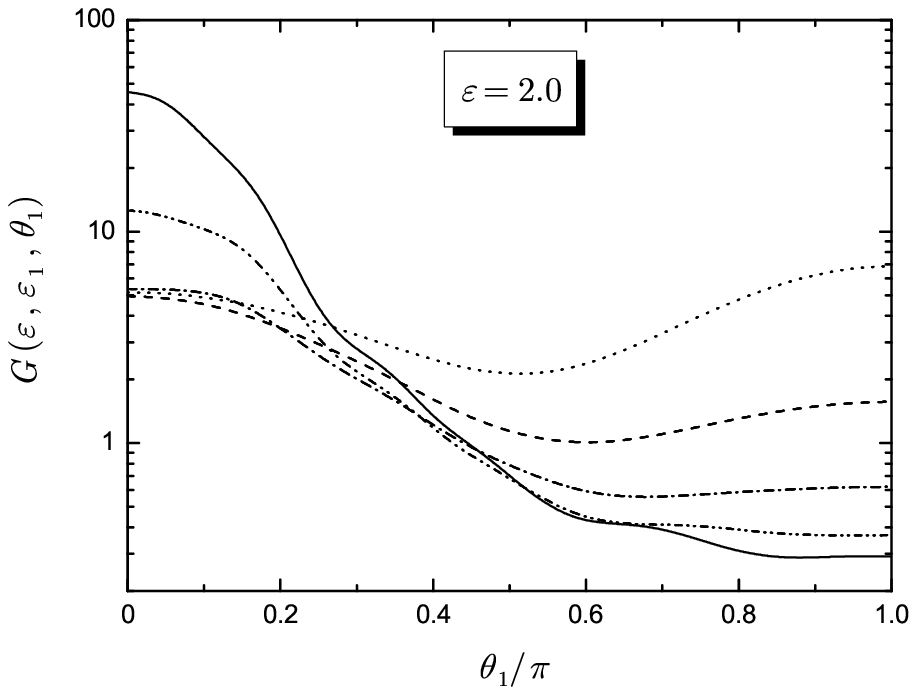}
\end{minipage}
\begin{minipage}[b]{0.49\textwidth}
\centering\includegraphics[width=0.9\textwidth,angle=0,clip]{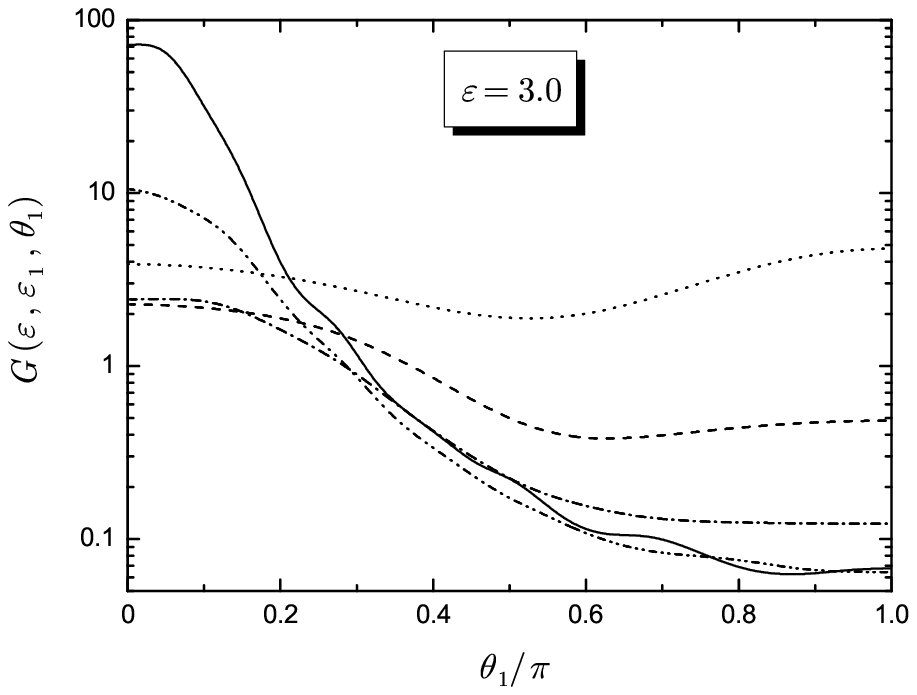}
\end{minipage}
\begin{minipage}[b]{0.49\textwidth}
\centering\includegraphics[width=0.9\textwidth,angle=0,clip]{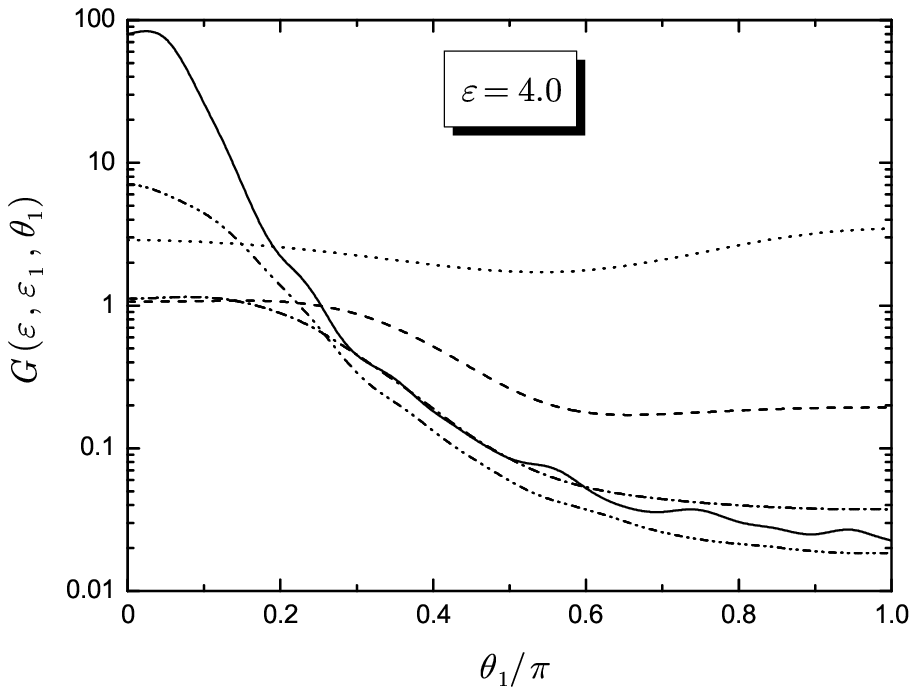}
\end{minipage}
\caption{\label{fig3} The universal function \eqref{eq30} is
calculated for different energies of incident and outgoing
electrons: dotted line, $\varepsilon_1=0$; dashed line,
$\varepsilon_1=0.25 (\varepsilon-1)$; dash-dotted line,
$\varepsilon_1=0.5(\varepsilon-1)$; dash-dot-dotted line,
$\varepsilon_1=0.75(\varepsilon-1)$; solid line,
$\varepsilon_1=\varepsilon-1$.}
\end{figure}

In Figs.~\ref{fig2} and \ref{fig3}, the universal functions
$G(\varepsilon,\varepsilon_1,\theta_1)$ are calculated for different
energies $\varepsilon$ of incident electrons. One can observe
several qualitative features in behavior of the universal curves
within the near-threshold energy domain. For very slow collisions,
the backward ($\theta_1 \simeq \pi$) electron scattering is more
probable than the forward ($\theta_1 \simeq 0$) scattering. In
particular, for $\varepsilon \lesssim 1.1$, this occurs for both
slow [$\varepsilon_1 \leqslant (\varepsilon-1)/2$] and fast
[$\varepsilon_1 \geqslant (\varepsilon-1)/2$] electrons. For
$\varepsilon \simeq 1.2$, the cross sections for backward and
forward scattering become to be of the comparable magnitude. The
slowest electrons ($\varepsilon_1 \simeq 0$) are scattered
predominantly backward, while the fastest electrons ($\varepsilon_1
\simeq \varepsilon -1$) are scattered mainly forward. With
increasing incident energies ($\varepsilon \geqslant 1.5$), the
dominant contribution to the ionization cross section arises from
the fast outgoing electrons, which are scattered forward at small
angles $\theta_1$. The backward scattering turns out to be
increasingly suppressed. In addition, the angular distribution of
the slow electrons becomes to be more isotropic.

Integrating Eq.~\eqref{eq29} over the solid angle $d\Omega_1$ yields
the energy distribution for outgoing electrons
\begin{equation}
\frac{d\sigma^+_{\mathrm{K}}}{d\varepsilon_1} =
\frac{\sigma_0}{Z^4} F(\varepsilon,\varepsilon_1) \, , \label{eq33}\\
\end{equation}
where $F(\varepsilon,\varepsilon_1)$ is given by Eq.~\eqref{eq31}.
This universal function has been already studied in the entire
non-relativistic energy domain \cite{7}.

\subsection{Positron impact}

Let us now consider the ionization of a K-shell bound electron due
to the inelastic positron scattering. As in the case of the electron
impact, the incident positron can be characterized by the energy
$E=\bm{p}^2/(2 m)$ and the asymptotic momentum $\bm{p}$, while the
scattered positron is characterized by the energy $E_1=\bm{p}_1^2/(2
m)$ and the momentum $\bm{p}_1$. The energy-conservation law keeps
the same form, namely, $E -I =E_1 + E_2$, where $E_2=\bm{p}_2^2/(2
m)$ denotes the energy of ejected electron. Since the interacting
particles are not identical, the exchange effect is absent.
Accordingly, the ionization process is represented by the diagram
depicted in Fig.~\ref{fig1}(a) only. The differential cross section,
which describes the universal energy and angular distributions for
outgoing positrons, is given by the same formulas
\eqref{eq29}--\eqref{eq32}, where the function
$T^{l'l_1'}_{ll_1l_2}$ contains only the first term on the
right-hand side of Eqs.~\eqref{eq27} and \eqref{eq31}. In
Eqs.~\eqref{eq8}--\eqref{eq10}, which correspond to the Coulomb wave
functions of the incident and scattered positrons, one needs to
employ the case of repulsive field of the atomic nucleus (upper
signs). In particular, the phase shift $\Delta^{l'l_1'}_{ll_1}$
reads now as follows $\Delta^{l'l_1'}_{ll_1}= \Delta_{ll_1}-
\Delta_{l'l'_1}$, where $\Delta_{ll_1}=
\delta^{(+)}_{pl}+\delta^{(+)}_{p_1l_1}+ \pi(l-l_1)/2$.

\begin{figure}[tbhp]
\begin{minipage}[t]{0.49\textwidth}
\centering\includegraphics[width=0.9\textwidth,angle=0,clip]{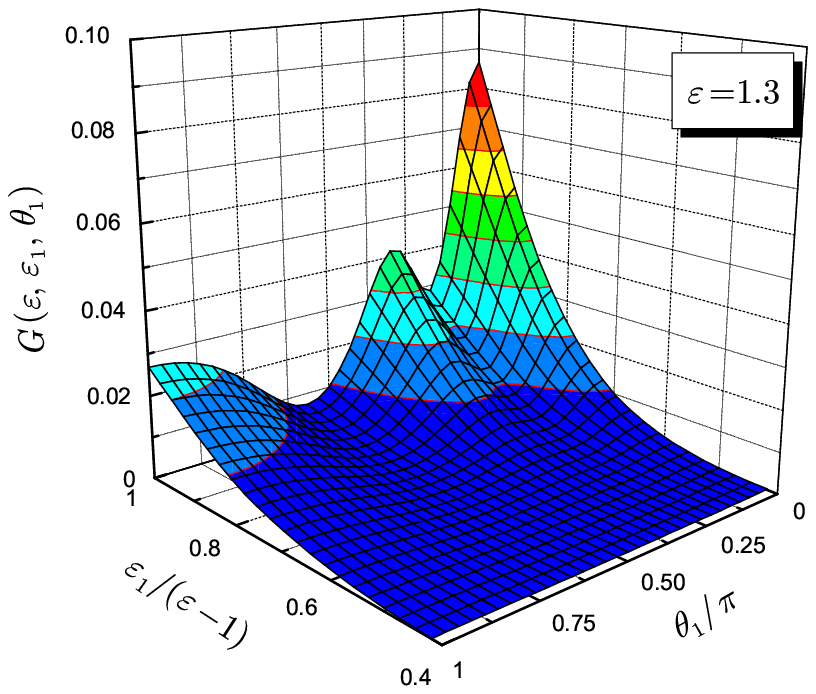}
\end{minipage}
\begin{minipage}[t]{0.49\textwidth}
\centering\includegraphics[width=0.9\textwidth,angle=0,clip]{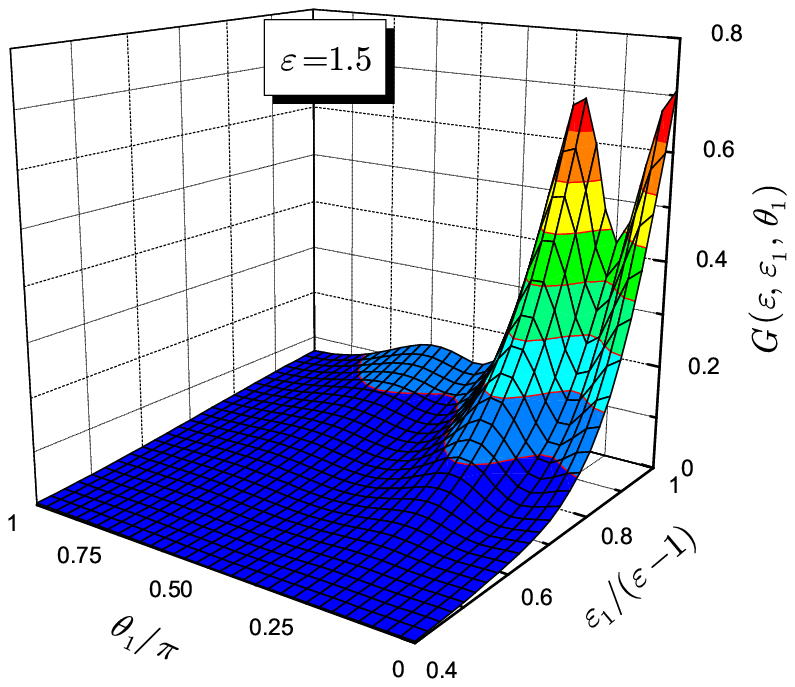}
\end{minipage}
\begin{minipage}[h]{0.49\textwidth}
\centering\includegraphics[width=0.9\textwidth,angle=0,clip]{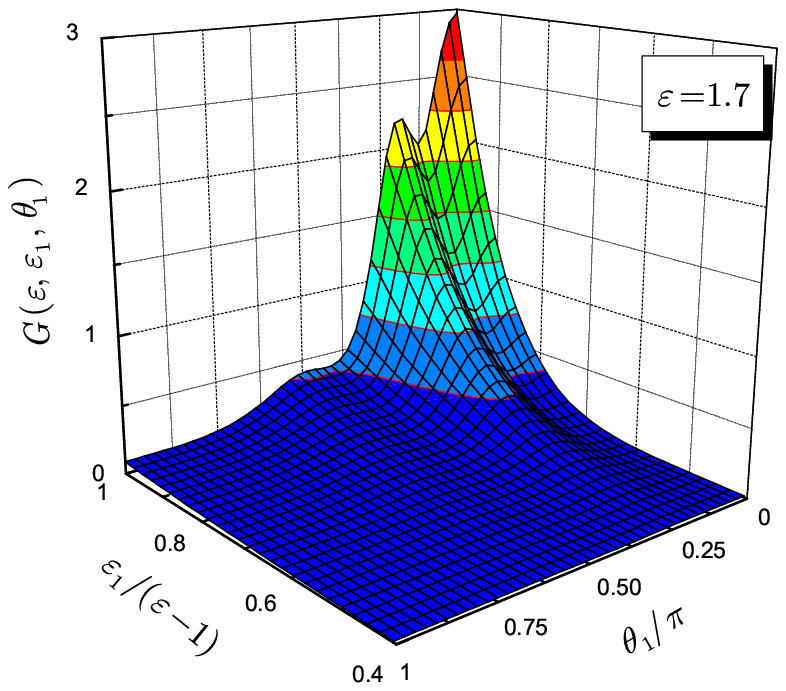}
\end{minipage}
\begin{minipage}[h]{0.49\textwidth}
\centering\includegraphics[width=0.9\textwidth,angle=0,clip]{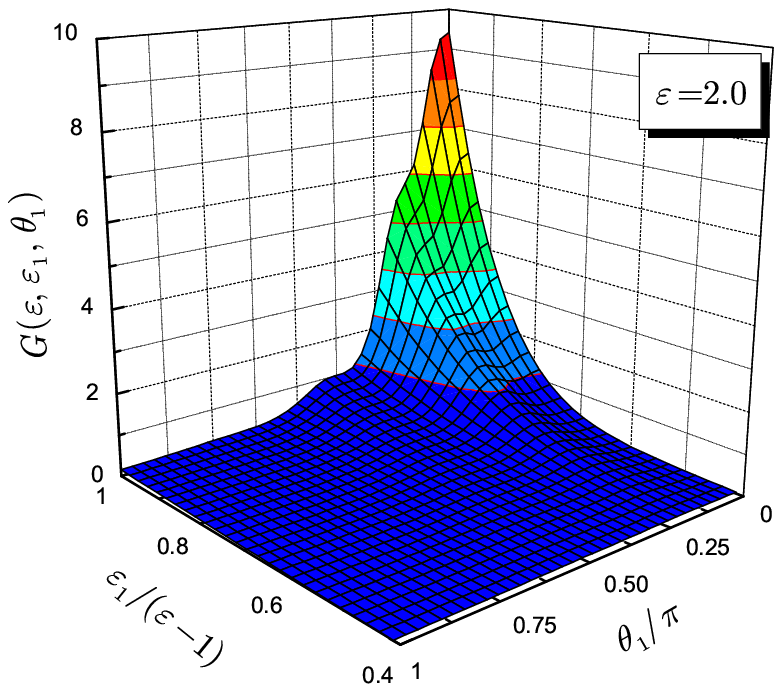}
\end{minipage}
\begin{minipage}[b]{0.49\textwidth}
\centering\includegraphics[width=0.9\textwidth,angle=0,clip]{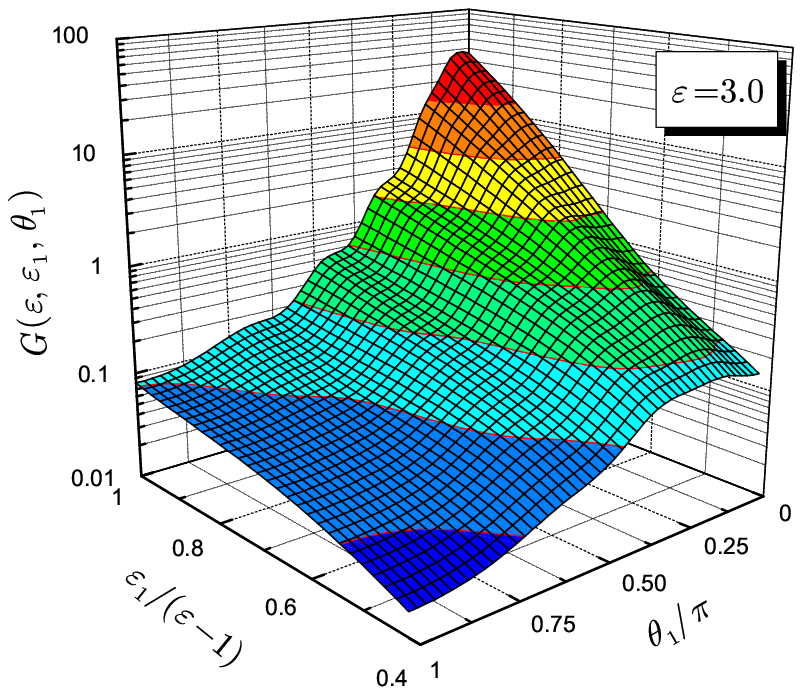}
\end{minipage}
\begin{minipage}[b]{0.49\textwidth}
\centering\includegraphics[width=0.9\textwidth,angle=0,clip]{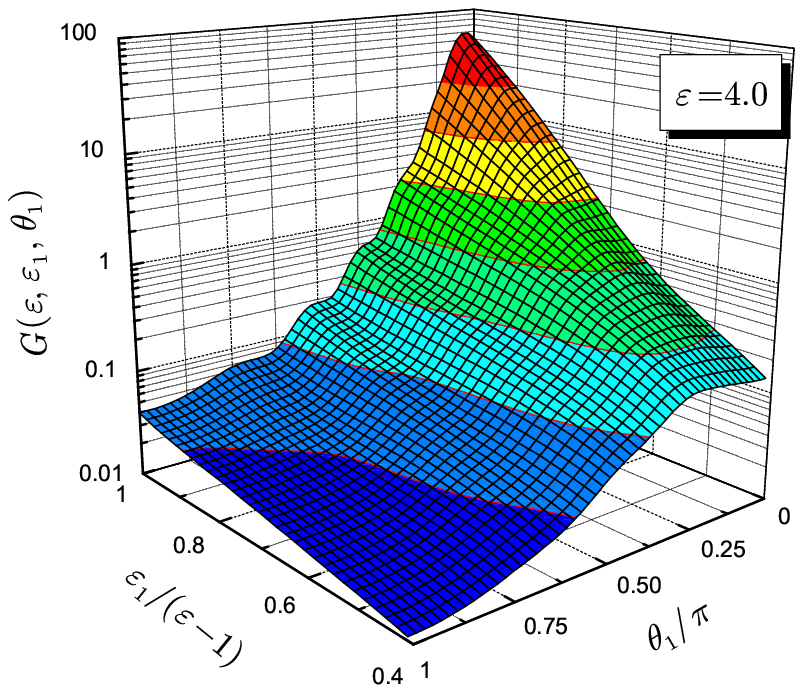}
\end{minipage}
\caption{\label{fig4} The universal function \eqref{eq30} is
calculated for different energies $\varepsilon$ of incident
positrons. The variable $\varepsilon_1$ is the energy of outgoing
positron, which is scattered at the angle $\theta_1$ with respect to
direction of the asymptotic momentum $\bm{p}$.}
\end{figure}

\begin{figure}[tbhp]
\begin{minipage}[t]{0.49\textwidth}
\centering\includegraphics[width=0.9\textwidth,angle=0,clip]{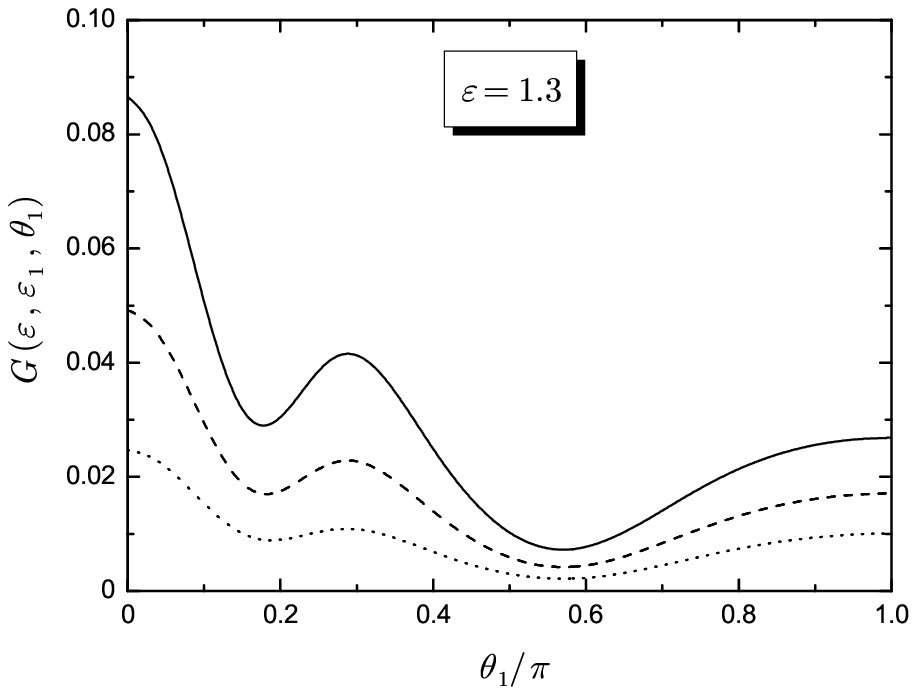}
\end{minipage}
\begin{minipage}[t]{0.49\textwidth}
\centering\includegraphics[width=0.9\textwidth,angle=0,clip]{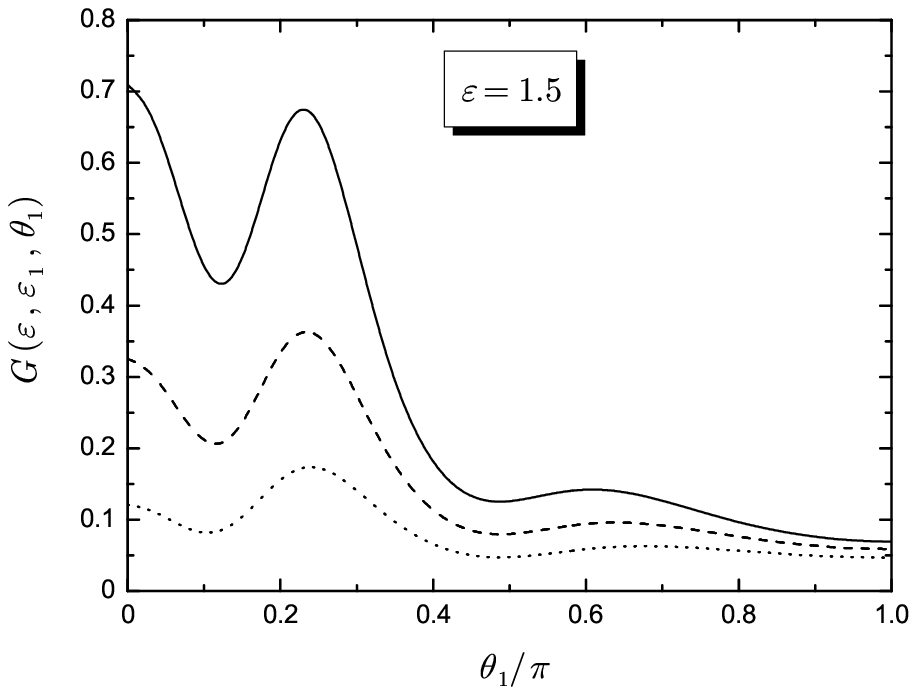}
\end{minipage}
\begin{minipage}[h]{0.49\textwidth}
\centering\includegraphics[width=0.9\textwidth,angle=0,clip]{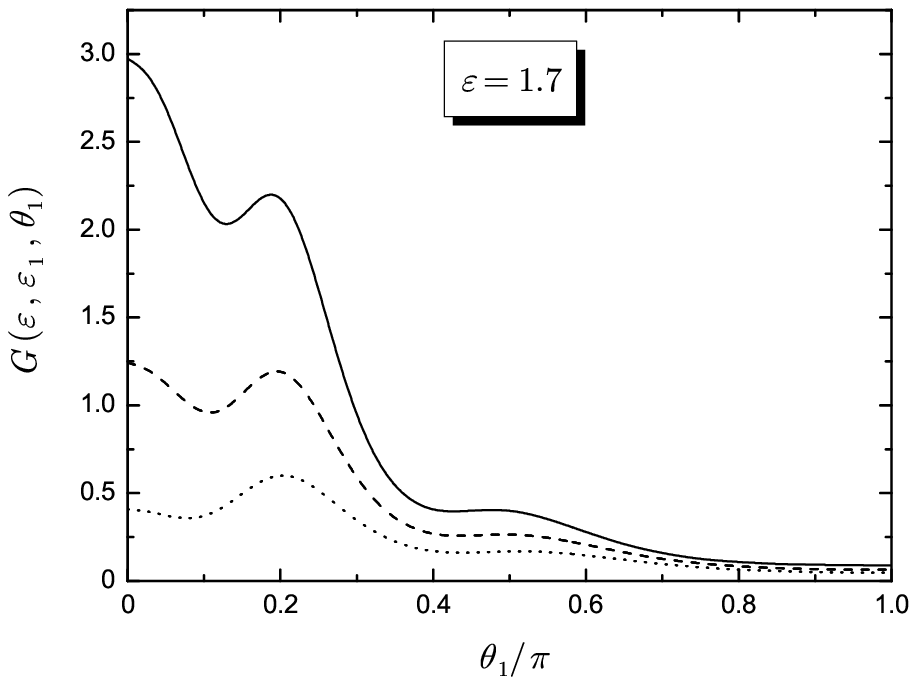}
\end{minipage}
\begin{minipage}[h]{0.49\textwidth}
\centering\includegraphics[width=0.9\textwidth,angle=0,clip]{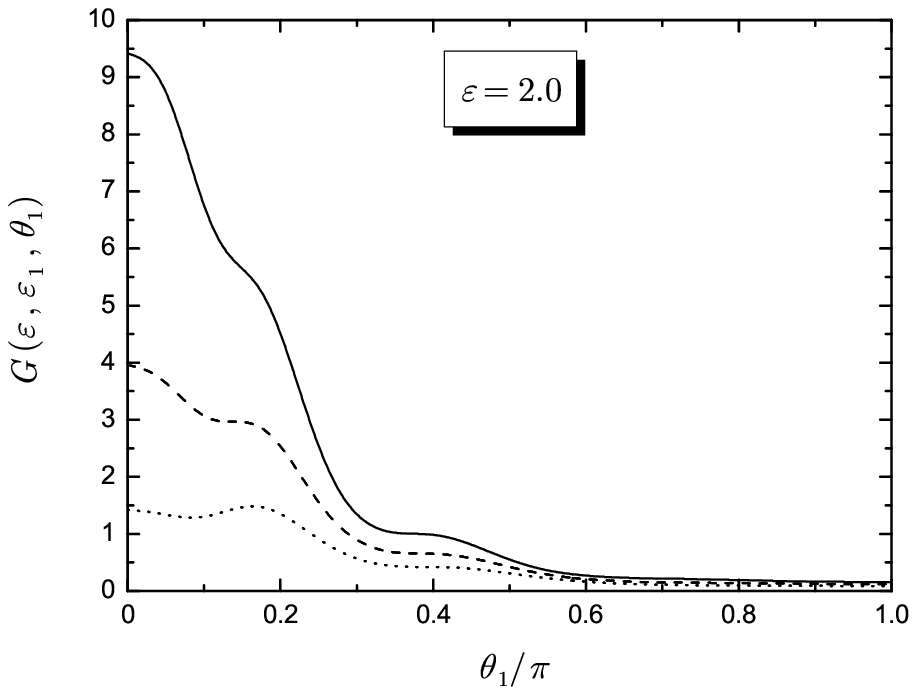}
\end{minipage}
\begin{minipage}[b]{0.49\textwidth}
\centering\includegraphics[width=0.9\textwidth,angle=0,clip]{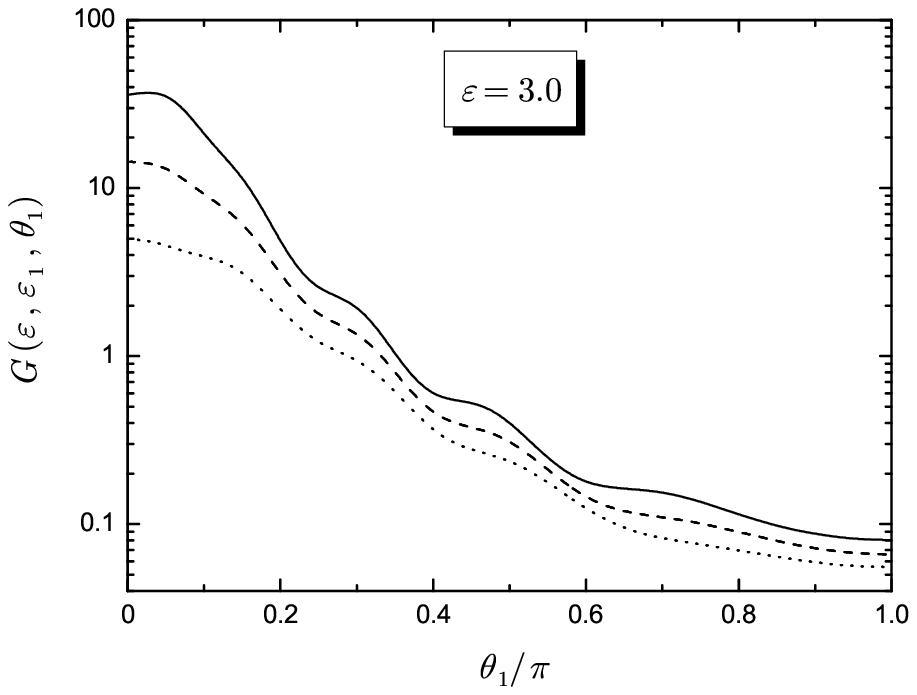}
\end{minipage}
\begin{minipage}[b]{0.49\textwidth}
\centering\includegraphics[width=0.9\textwidth,angle=0,clip]{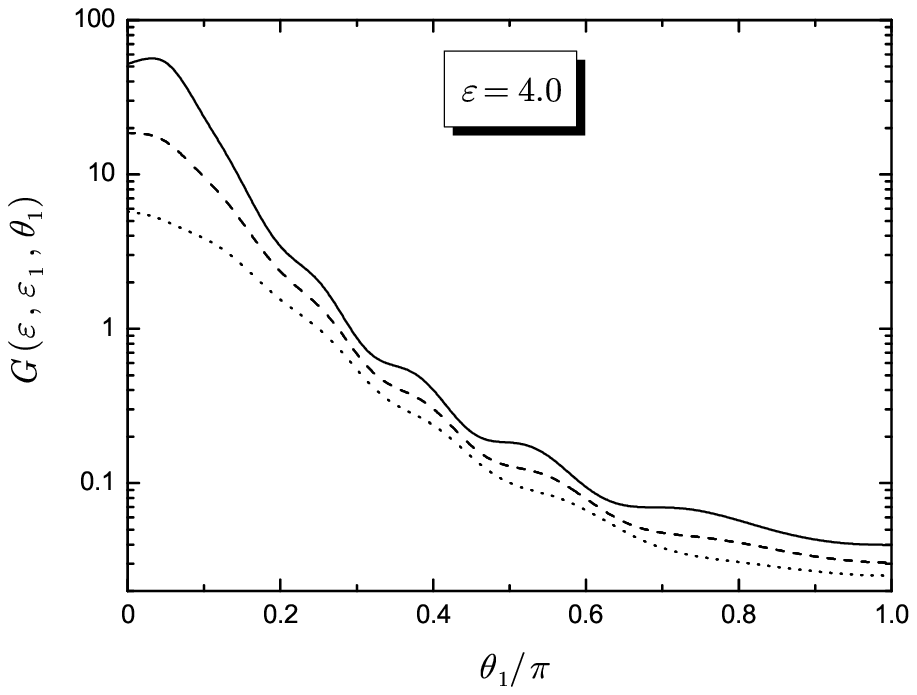}
\end{minipage}
\caption{\label{fig5} The universal function \eqref{eq30} is
calculated for different energies of incident and outgoing
positrons: dotted line, $\varepsilon_1=0.8 (\varepsilon-1)$; dashed
line, $\varepsilon_1=0.9(\varepsilon-1)$; solid line,
$\varepsilon_1=\varepsilon-1 $. }
\end{figure}

In Figs.~\ref{fig4} and \ref{fig5}, the universal energy and angular
distributions for outgoing positrons are calculated for few values
of the dimensionless energy $\varepsilon$. Although the scattered
positron can have any energy within the range $0 \leqslant
\varepsilon_1 \leqslant \varepsilon -1$, the dominant contribution
to the ionization cross section arises from the fast positrons with
the energies $\varepsilon_1$, which are close enough to the excess
energy $\varepsilon -1$. For slow collisions with $\varepsilon
\lesssim 1.7$, the differential cross section contains three
pronounced maximums at different scattering angles $\theta_1$. With
increasing the incident energies up to $\varepsilon \sim 2$, the
maximums coalesce near the zeroth angle, so that the positrons are
preferably scattered in the forward cone ($\theta_1 < \pi/2$). With
further increasing the incident energies $\varepsilon$, the angular
distribution of the fast outgoing positrons exhibits weak
interference oscillations with increasing frequency.

The differential cross section, which describes the energy and
angular distributions of the electrons ejected by positron impact,
is given by
\begin{eqnarray}
\frac{d\sigma^+_{\mathrm{K}}}{d\varepsilon_2 d\Omega_2} &=&
\frac{\sigma_0}{4\pi Z^4}G(\varepsilon,\varepsilon_2,\theta_2) \, ,
\label{eq34}\\
G(\varepsilon,\varepsilon_2,\theta_2)&=&
F(\varepsilon,\varepsilon_2) + \sum_{L\geqslant
1}F_L(\varepsilon,\varepsilon_2) P_L(\cos \theta_2)
\,  , \label{eq35}\\
F(\varepsilon,\varepsilon_2) &=&\frac{1}{\varepsilon} \sum_{l, l_1,
l_2} T^{ll_2}_{ll_1 l_2}= \frac{1}{\varepsilon}\sum_{l, l_1,l_2}
\bigl(W^{l}_{l_1 l_2}\bigr)^2  \,  , \label{eq36}\\
F_L(\varepsilon,\varepsilon_2) &=& \frac{1}{\varepsilon} \sum_{l,
l_1,l_2}\sum_{l',l_2'} (-1)^{l_1}\cos \bigl(\Delta^{l'l_2'}_{ll_2}
\bigr)T^{l'l_2'}_{ll_1l_2} \Pi_{ll^{\phantom{*}}_2l'l_2'} C^{L0}_{l
0 l'0}C^{L0}_{l^{\phantom{*}}_2 0 l_2'0} \Biggl\{
\begin{matrix} l_2 & l_1 & l \\ l' &L &l_2' \end{matrix}
\Biggr\} \,  . \label{eq37}\\
T^{l'l_2'}_{ll_1l_2}&=& W^{l}_{l^{\vphantom{'}}_1l^{\vphantom{'}}_2}
W^{l'}_{l^{\vphantom{'}}_1l_2'}\,  , \qquad W^{l}_{l_1l_2} =
\frac{1}{\sqrt{\pi k k_1 k_2}} \frac{\Pi_{l_1}}{\Pi_{l_2}}
C^{l0}_{l_1 0 l_2 0} I^l_{l_1l_2}\, , \label{eq38}\\
I^l_{l_1l_2}&=& \int\limits_0^\infty dx^{\phantom{*}}_1 x_1^2
R^{(+)}_{k_1l_1}(x_1) R^{(+)}_{kl}(x_1) \int\limits_0^\infty
dx^{\phantom{*}}_2 x_2^2 R^{(-)}_{k_2l_2}(x_2)
\frac{x^{l_2}_<}{x^{l_2+1}_>} e^{-x_2} \,  . \label{eq39}
\end{eqnarray}
Here $\sigma_0=\pi a_0^2$, $a_0=1/(m\alpha)$, and
$\Delta^{l'l_2'}_{ll_2}=\Delta_{ll_2}-\Delta_{l'l'_2}$, where
$\Delta_{ll_2}= \delta^{(+)}_{pl}+\delta^{(-)}_{p_2l_2}+
\pi(l-l_2)/2$. The solid angle reads $d\Omega_2= 2\pi \sin \theta_2
d\theta_2$, where the ejection angle $\theta_2$ is enclosed by the
asymptotic momenta $\bm{p}$ and $\bm{p}_2$. The function
\eqref{eq37} is reduced to the function \eqref{eq36}, if $L=0$. The
energy $\varepsilon_2$ of ejected electrons lies within the range $0
\leqslant \varepsilon_2 \leqslant \varepsilon -1$.

\begin{figure}[tbhp]
\begin{minipage}[t]{0.49\textwidth}
\centering\includegraphics[width=0.9\textwidth,angle=0,clip]{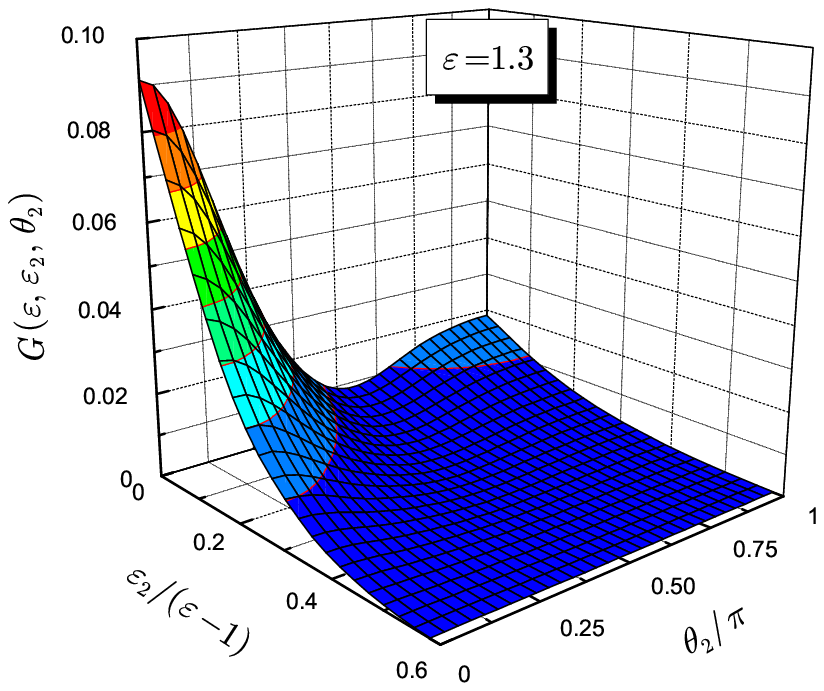}
\end{minipage}
\begin{minipage}[t]{0.49\textwidth}
\centering\includegraphics[width=0.9\textwidth,angle=0,clip]{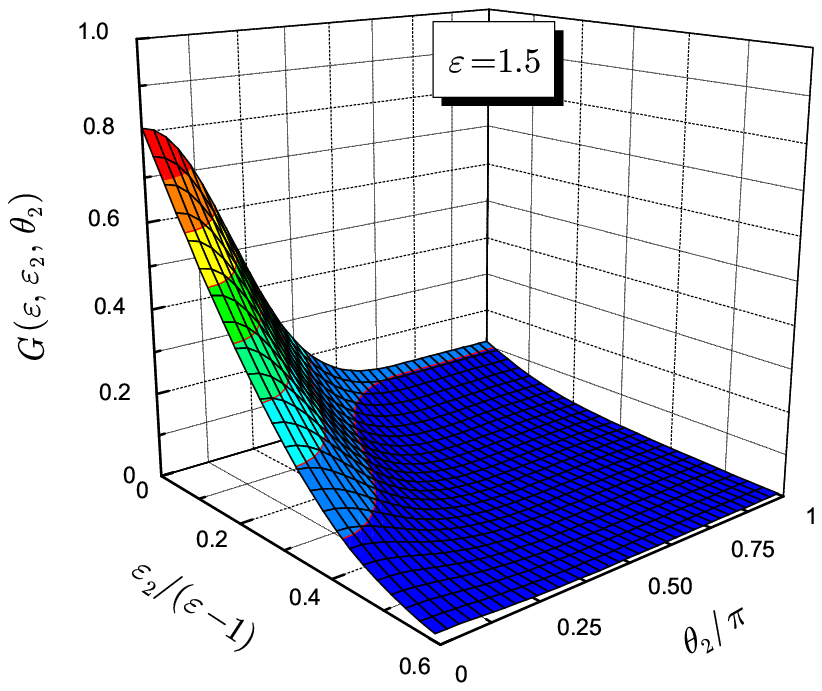}
\end{minipage}
\begin{minipage}[h]{0.49\textwidth}
\centering\includegraphics[width=0.9\textwidth,angle=0,clip]{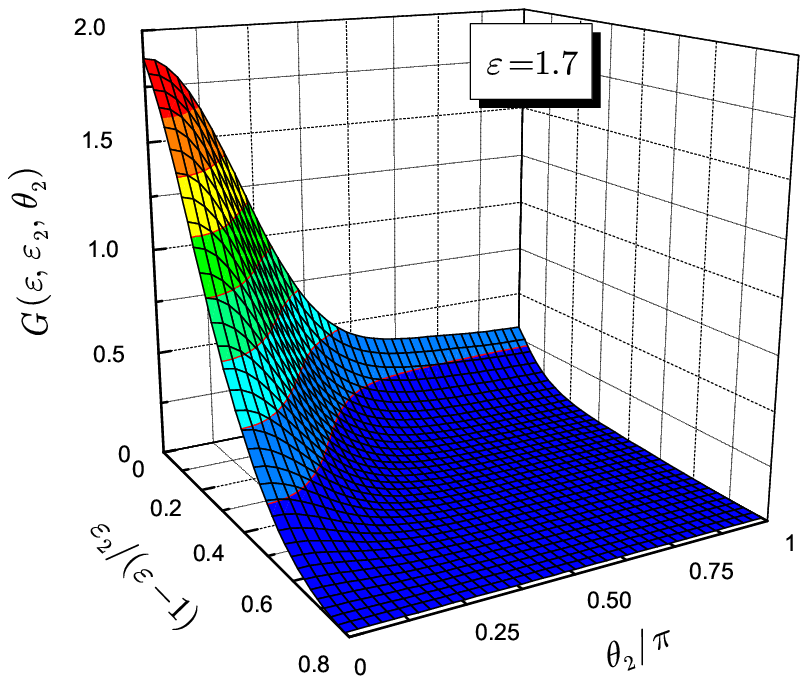}
\end{minipage}
\begin{minipage}[h]{0.49\textwidth}
\centering\includegraphics[width=0.9\textwidth,angle=0,clip]{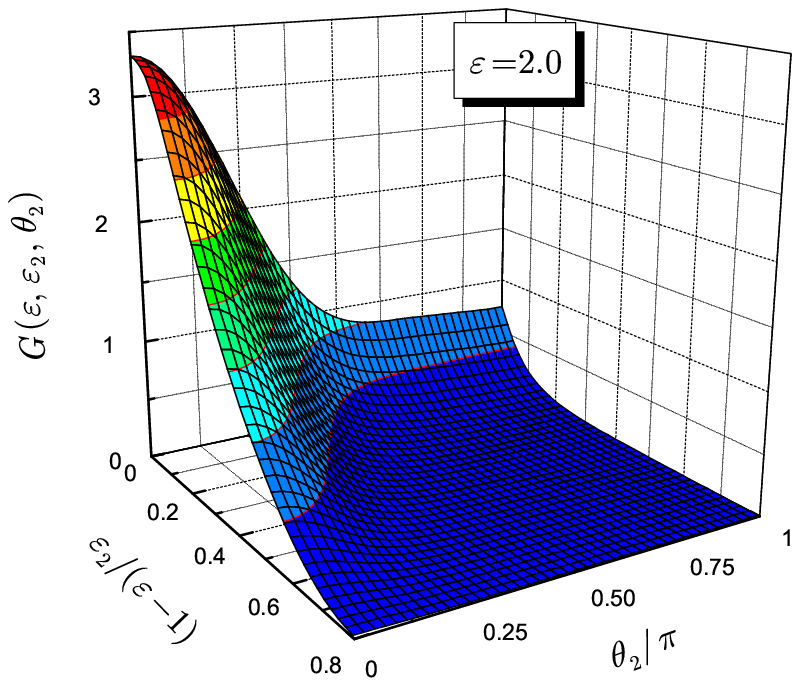}
\end{minipage}
\begin{minipage}[b]{0.49\textwidth}
\centering\includegraphics[width=0.9\textwidth,angle=0,clip]{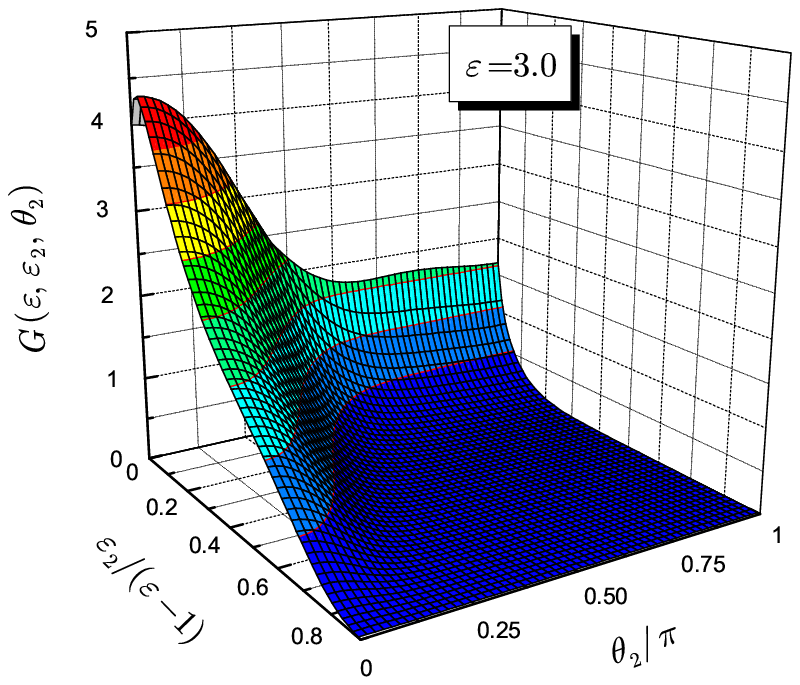}
\end{minipage}
\begin{minipage}[b]{0.49\textwidth}
\centering\includegraphics[width=0.9\textwidth,angle=0,clip]{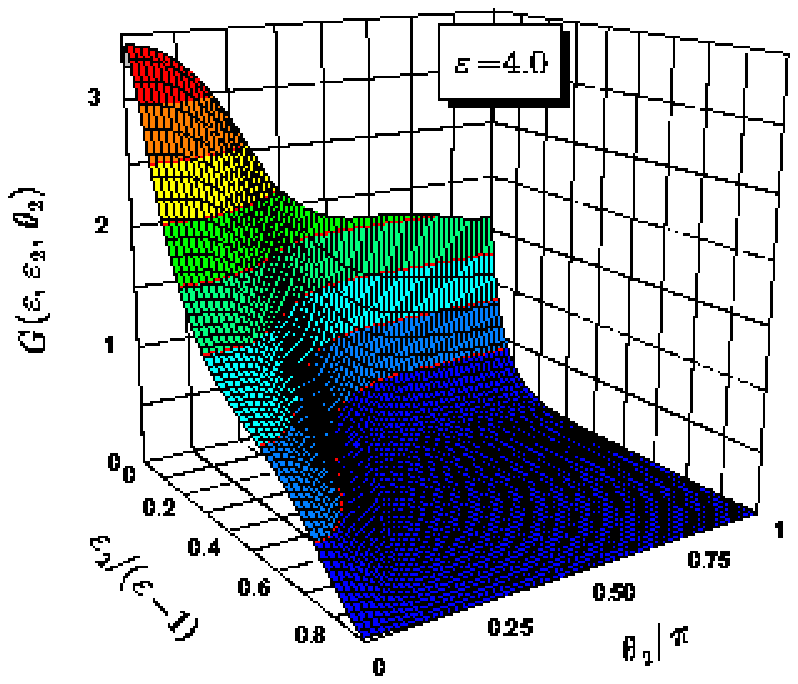}
\end{minipage}
\caption{\label{fig6} The universal function \eqref{eq35} is
calculated for different energies $\varepsilon$ of incident
positrons. The variable $\varepsilon_2$ is the energy of outgoing
electron, which is ejected at the angle $\theta_2$ with respect to
direction of the asymptotic momentum $\bm{p}$.}
\end{figure}

\begin{figure}[tbhp]
\begin{minipage}[t]{0.49\textwidth}
\centering\includegraphics[width=0.9\textwidth,angle=0,clip]{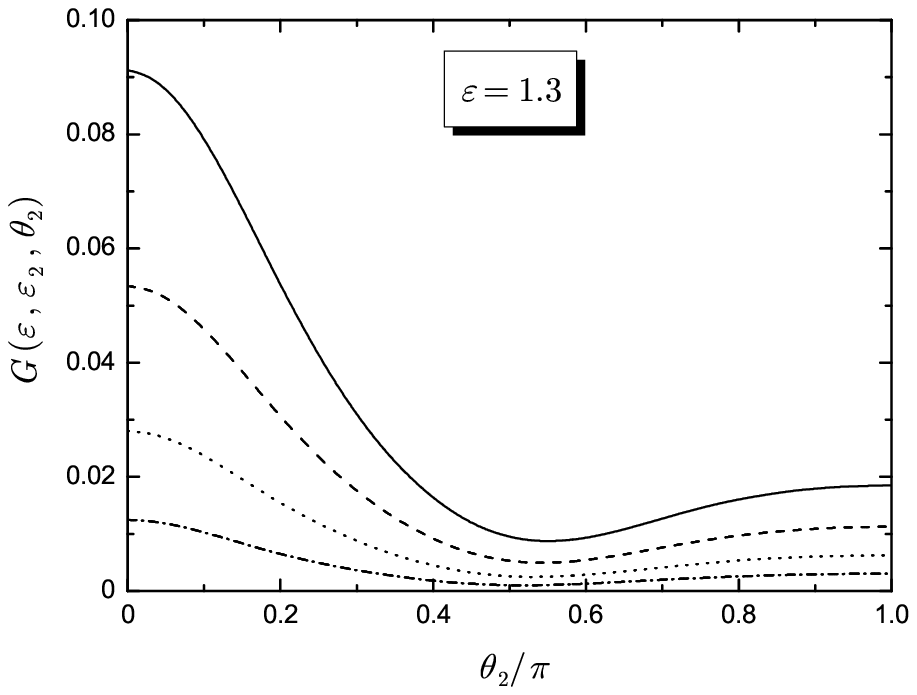}
\end{minipage}
\begin{minipage}[t]{0.49\textwidth}
\centering\includegraphics[width=0.9\textwidth,angle=0,clip]{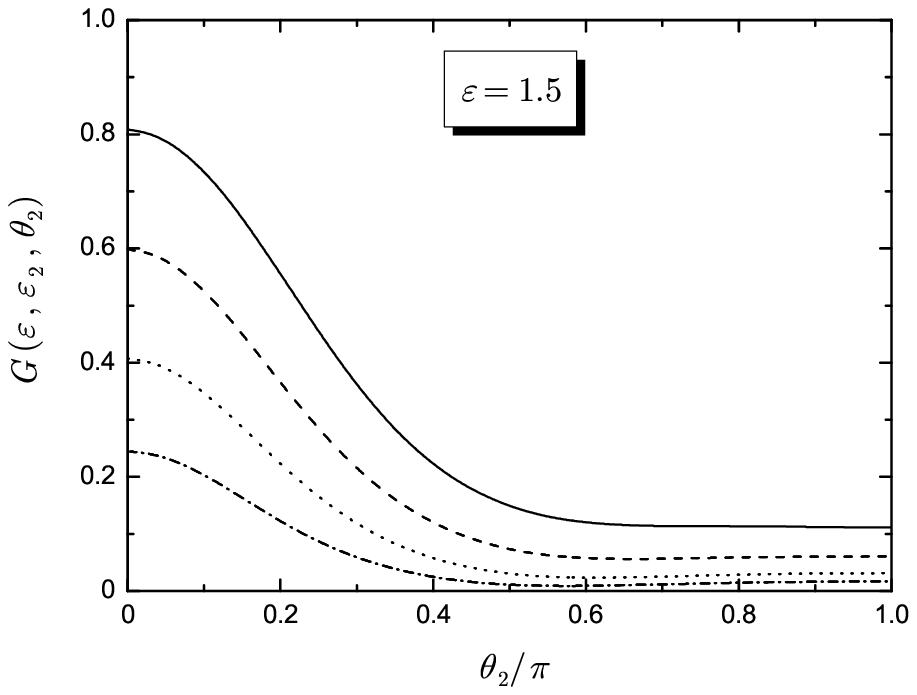}
\end{minipage}
\begin{minipage}[h]{0.49\textwidth}
\centering\includegraphics[width=0.9\textwidth,angle=0,clip]{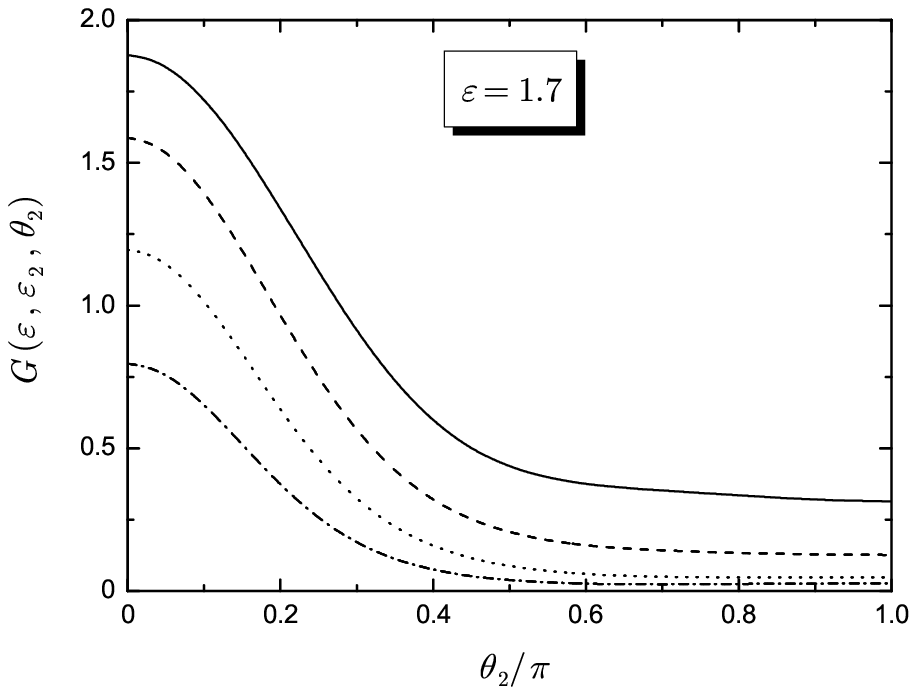}
\end{minipage}
\begin{minipage}[h]{0.49\textwidth}
\centering\includegraphics[width=0.9\textwidth,angle=0,clip]{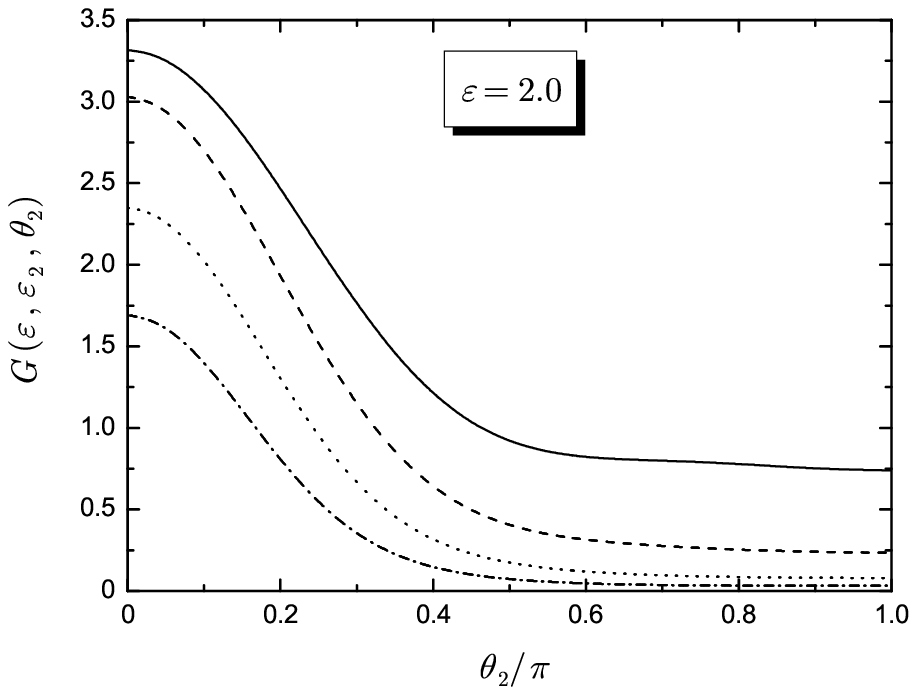}
\end{minipage}
\begin{minipage}[b]{0.49\textwidth}
\centering\includegraphics[width=0.9\textwidth,angle=0,clip]{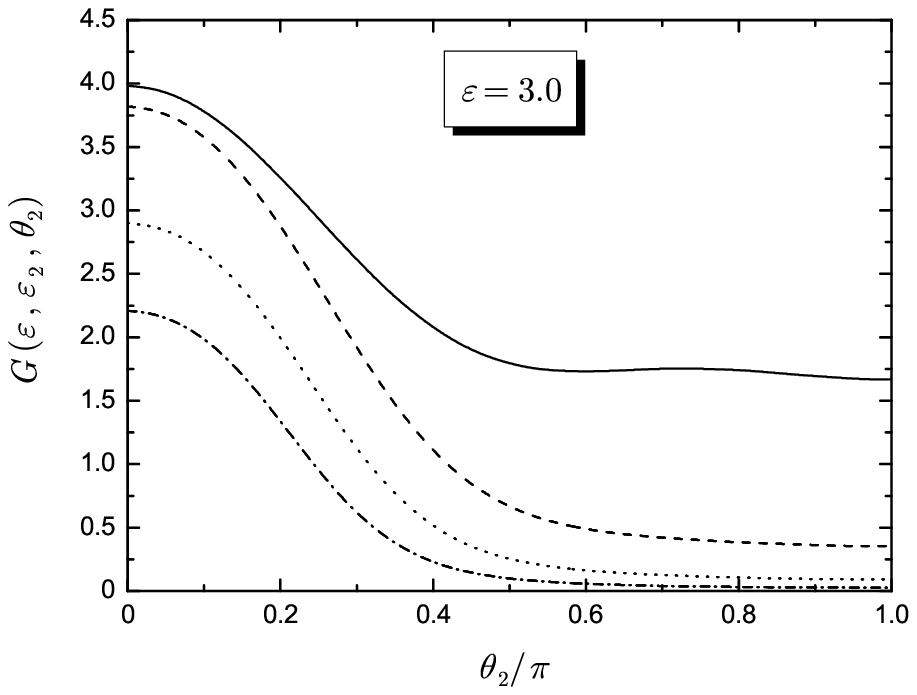}
\end{minipage}
\begin{minipage}[b]{0.49\textwidth}
\centering\includegraphics[width=0.9\textwidth,angle=0,clip]{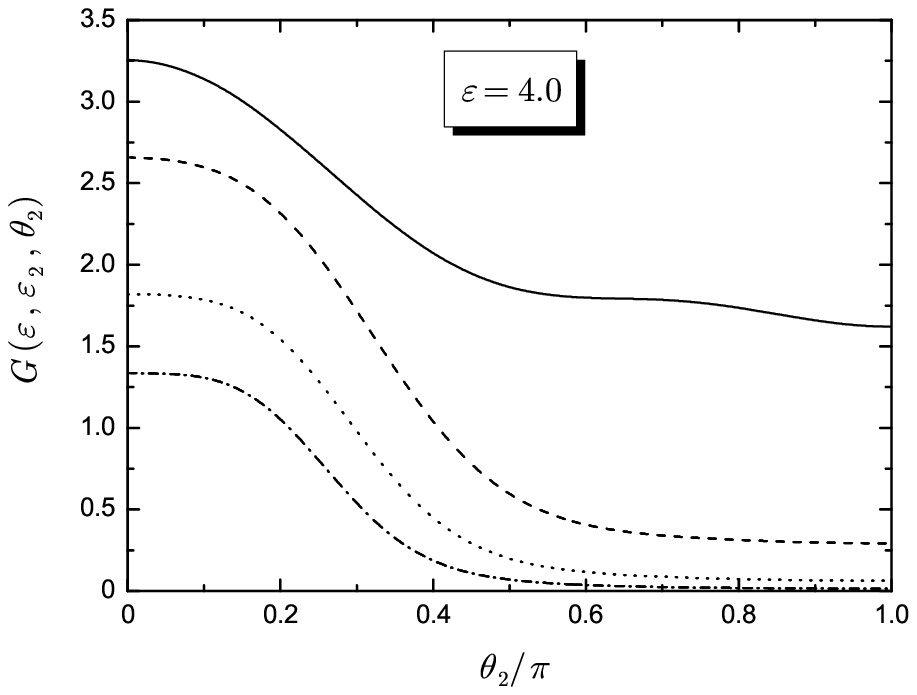}
\end{minipage}
\caption{\label{fig7} The universal function \eqref{eq35} is
calculated for different energies of incident positrons and outgoing
electrons: dash-dotted line, $\varepsilon_2=0.3 (\varepsilon-1)$;
dotted line, $\varepsilon_2=0.2 (\varepsilon-1)$; dashed line,
$\varepsilon_2=0.1(\varepsilon-1)$; solid line, $\varepsilon_2=0 $.}
\end{figure}

In Figs.~\ref{fig6} and \ref{fig7}, the universal function
\eqref{eq35} is calculated for different values of the dimensionless
energy $\varepsilon$ of incident positrons. Within the
near-threshold energy domain, the electron emission occurs
preferably at small angles $\theta_2 \simeq 0$, although for
$\varepsilon \lesssim 1.3$, the differential cross section exhibits
also a weak maximum at $\theta_2=\pi$. The total cross section is
exhausted within the range of small energies $\varepsilon_2$ and
small angles $\theta_2$. With increasing the incident energies
$\varepsilon$, the relative amount of slow electrons ejected at
arbitrary angles $\theta_2$ is growing.

Integrating Eq.~\eqref{eq34} over the solid angle $d\Omega_2$ yields
the energy distribution for outgoing electrons
\begin{equation}
\frac{d\sigma^+_{\mathrm{K}}}{d\varepsilon_2} =
\frac{\sigma_0}{Z^4} F(\varepsilon,\varepsilon_2) \, , \label{eq40}\\
\end{equation}
where $F(\varepsilon,\varepsilon_2)$ is given by Eq.~\eqref{eq36}.
This universal function has been already studied in the entire
non-relativistic energy domain \cite{8}. Note also that, for any
incident positron energy $\varepsilon$, the universal function
$F(\varepsilon,\varepsilon_1)$, which describes the energy
distribution for scattered positrons, is symmetrical to the function
$F(\varepsilon,\varepsilon_2)$ with respect to the vertical axis
crossing the energy interval in the middle point
$\varepsilon_1=\varepsilon_2=(\varepsilon -1)/2$.

\section{Generalization to arbitrary atomic target and conclusions}

Equations~\eqref{eq29}, \eqref{eq33}, \eqref{eq34}, and \eqref{eq40}
describe the single ionization of hydrogen-like ions in the ground
state. However, due to universality of the scaling behavior, these
formulas can be easily generalized on the case of arbitrary
non-relativistic atomic targets, in which the K shell is completely
occupied. Firstly, the ionization cross sections should be
multiplied by a factor $2$, taking into account the number of
K-shell electrons. Secondly, one needs to simulate the screening
effect of the passive electrons on the active K-shell electron,
participating in the ionization process. This can be achieved by
substitution of the true nuclear charge $Z$ by the corresponding
effective value $Z_{\textrm{eff}}$, which is defined via \cite{12}
\begin{equation}\label{eq41}
I_{\textrm{exp}} =\frac{ m }{2}(\alpha Z_{\textrm{eff}})^2 \,  ,
\end{equation}
where $I_{\textrm{exp}}$ is the experimentally observable threshold
for the single K-shell ionization. Accordingly, the energies of
incident and outgoing particles should be calibrated in units of the
experimental value $I_{\mathrm{exp}}$, that is, $\varepsilon
=E/I_{\mathrm{exp}}$ and $\varepsilon_i=E_i/I_{\mathrm{exp}}$,
$(i=1,2)$. The universal functions \eqref{eq30} and \eqref{eq35}
depicted in Figs.~\ref{fig2}--\ref{fig7} keep the same scaling
behavior for non-relativistic atomic targets with arbitrary nuclear
charge $Z \gg 1$.

Concluding, we have deduced the universal scaling behavior of
differential cross sections for the single K-shell ionization by
electron and positron impact. The results are obtained within the
framework of non-relativistic perturbation theory, taking into
account the one-photon exchange diagrams. The universal scaling laws
can be applied for both multicharged ions and neutral atoms with
moderate values of the nuclear charge number $Z$. The interference
oscillations in doubly differential cross sections for inelastic
scattering of low-energy positrons deserve further experimental
verification.

\section*{Acknowledgments}

AM and AN are grateful to the Dresden University of Technology for
hospitality and for financial support from Max Planck Institute for
the Physics of Complex Systems. This research was financed in part
by RFBR under Grant No. 08-02-00460-a and by GSI.

\newpage

\begin{thebibliography}{20}
\bibitem{10} The use of the same notations for the electron mass and projection
of the orbital angular momentum is not confusing, because the latter
is related with the angular dependence of wave functions only.
\end{thebibliography}
\end{document}